\documentclass[pre,aps,10pt,superscriptaddress,twocolumn,floatfix]{revtex4-1}
\usepackage{graphicx,color}
\graphicspath{{figpdf_paper/}, {figpdf_paper_preliminary/}}
\usepackage[nice]{nicefrac}							% Nice fractions 1/2
\usepackage{amsmath,amssymb,bm}
\usepackage{braket}
\usepackage[plainpages=false,pdfpagelabels,colorlinks=true,linkcolor=red,urlcolor=blue,citecolor=blue,pdftitle={},pdfauthor={},pdfdisplaydoctitle=true,pdfduplex=DuplexFlipLongEdge]{hyperref}

\newcommand{\mc}{\mathcal}

\definecolor{darkred}{rgb}{0.90,0.2,0.2}
\definecolor{darkgreen}{rgb}{0,0.60,.2}
\definecolor{darkblue}{rgb}{0.1,0.3,1}
\definecolor{grey}{cmyk}{0,0,0,0.25}
\definecolor{orange}{cmyk}{0,0.6,0.8,0}

\begin{document}

\title{Similarity between a many-body quantum avalanche model \\ and the ultrametric random matrix model}

\author{Jan \v Suntajs}
\affiliation{Department of Theoretical Physics, J. Stefan Institute, SI-1000 Ljubljana, Slovenia}
\affiliation{Faculty of Mechanical Engineering, University of Ljubljana, SI-1000 Ljubljana, Slovenia\looseness=-1}
\author{Miroslav Hopjan}
\affiliation{Department of Theoretical Physics, J. Stefan Institute, SI-1000 Ljubljana, Slovenia}
\author{Wojciech De Roeck}
\affiliation{Institute of Theoretical Physics, K.U. Leuven, 3001 Leuven, Belgium}
\author{Lev Vidmar}
\affiliation{Department of Theoretical Physics, J. Stefan Institute, SI-1000 Ljubljana, Slovenia}
\affiliation{Department of Physics, Faculty of Mathematics and Physics, University of Ljubljana, SI-1000 Ljubljana, Slovenia\looseness=-1}

%\date{\today}

\begin{abstract}
In the field of ergodicity-breaking phases, it has been recognized that quantum avalanches can destabilize many-body localization at a wide range of disorder strengths. This has in particular been demonstrated by the numerical study of a toy model, sometimes simply called the ‘’avalanche model’’ or the ‘’quantum sun model’’ [\href{https://doi.org/10.1103/PhysRevLett.129.060602}{Phys.~Rev.~Lett.~{\bf 129},~060602~(2022)}], which consists of an ergodic seed coupled to a perfectly localized material.  In this paper, we connect this toy model to a well-studied model in random matrix theory, the ultrametric ensemble.   We conjecture that the models share the following features.  1) The location of the critical point may be predicted sharply by analytics. 2) On the localized site, both models exhibit Fock space localization. 3) There is a manifold of critical points. On the critical manifold, the eigenvectors exhibit nontrivial multifractal behaviour that can be tuned by moving on the manifold.  4) The spectral statistics at criticality is intermediate between Poisson statistics and random matrix statistics, also tunable on the critical manifold.   We confirm numerically these properties.
\end{abstract}

\maketitle

\section{Introduction}

Our knowledge of quantum thermalization in many-body systems has significantly improved in the last years~\cite{dalessio_kafri_16, mori_ikeda_18, deutsch_18, borgonovi_izrailev_16}.
A particularly exciting aspect of quantum thermalization is its ability to occur in perfectly isolated systems.
This is possible since the information about thermalization can be detected even on a level of Hamiltonian eigenstates~\cite{deutsch_91, srednicki_94, rigol_dunjko_08, dalessio_kafri_16}.
Valuable contribution to this development was provided by experimental advances to realize nearly perfectly isolated many-body quantum systems~\cite{kinoshita_wenger_06, gring_kuhnert_12, Trotzky2012, meinert13, langen15a, clos_porras_16, Kaufman2016, Neill2016, tang_kao_18}.

Of equal importance is to understand the boundaries of quantum thermalization and the conditions under which ergodicity breaking phase transitions may take place.
At the current stage of research, it is particularly important to establish toy models of ergodicity breaking phase transitions, which exhibit clear features of critical behavior already in finite systems amenable to numerical simulations.
Advances in this subject may bring new impetus for ongoing experimental activities~\cite{schreiber_hodgman_15, smith_lee_16, lueschen_bordia_17, rispoli_lukin_19, guo_cheng_21, gong_moraesneto_21, chiaro_neill_22, filho_izquierdo_22, leonard_kim_23}, as well as provide new perspective into many-body localization~\cite{basko_aleiner_06, gornyi_mirlin_05, oganesyan_huse_07, pal_huse_10}, for which different perspectives about its stability in the thermodynamic limit have recently been formulated~\cite{suntajs_bonca_20a, suntajs_bonca_20b, kieferemmanouilidis_unanyan_20, panda_scardicchio_20, sierant_delande_20, sierant_lewenstein_20, sels_polkovnikov_21, kieferemmanouilidis_unanyan_21, leblond_sels_21, vidmar_krajewski_21, abanin_bardarson_21, corps_molina_21, prakash_pixley_21, schliemann_costa_21, hopjan_orso_21, solorzano_santos_21, detomasi_khaymovich_21, krajewski_vidmar_22, crowley_chandran_22, ghosh_znidaric_22, bolther_kehrein_22, yintai_yufeng_22, sels_22, sierant_zakrzewski_22, morningstar_colmenarez_22, sutradhar_ghosh_22, trigueros_cheng_22, shi_khemani_22, sels_polkovnikov_23, peacock_sels_23, krajewski_vidmar_23, evers_bera_23}.

The theory of quantum avalanches provides a mechanism of thermalization in interacting systems in the absence of translational invariance, when small ergodic regions coexist with  mesoscopic nonergodic regions~\cite{deroeck_huveneers_17}. 
The theory explains why a seemingly stable nonergodic region, which exhibits a vanishingly small coupling to an ergodic region, may eventually thermalize~\cite{deroeck_huveneers_17, deroeck_imbrie_17, luitz_huveneers_17, thiery_huveneers_18, goihl_eisert_19, gopalakrishnan_huse_19, potirniche_banerjee_19, crowley_chandran_20, sels_22, morningstar_colmenarez_22, suntajs_vidmar_22, crowley_chandran_22b}.
At the same time, it also provides conditions for the breakdown of thermalization, which have been recently tested numerically in various quantum systems~\cite{luitz_huveneers_17, crowley_chandran_20, suntajs_vidmar_22, pawlik_sierant_23,hopjan2023,hopjan2023a}.

\begin{figure}[!t]
\centering
\includegraphics[width=0.90\columnwidth]{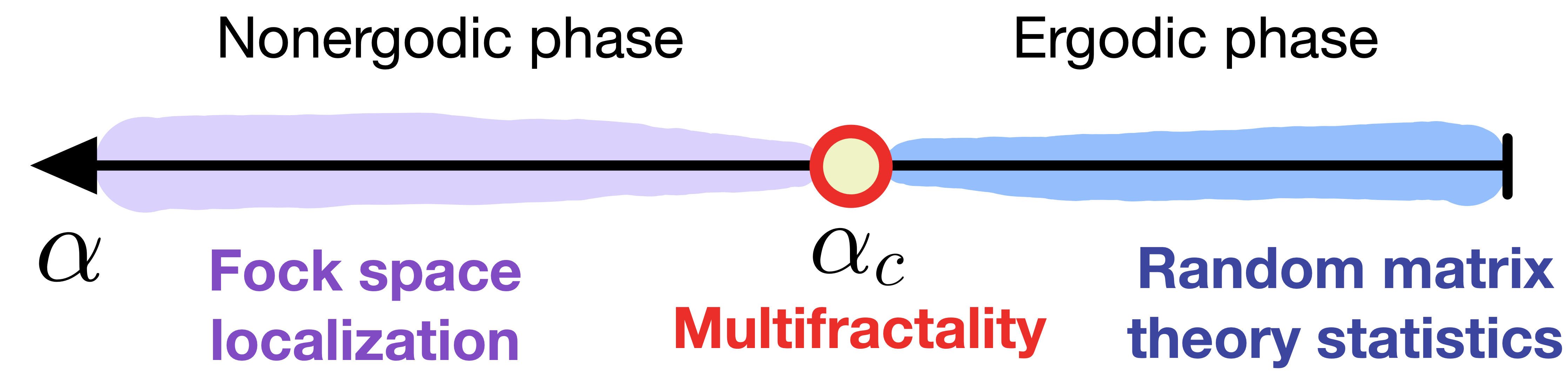}
\caption{
Phase diagram of the avalanche models studied in this work, as a function of the tuning parameter $\alpha$ that drives the system from an ergodic phase (in which short-range spectral statistics comply with random matrix theory predictions) to a nonergodic phase (that exhibits Fock space localization).
At the critical point, $\alpha=\alpha_c$, the spectral statistics may not comply with random matrix theory predictions and the Hamiltonian eigenstates exhibit multifractality in the basis of uncoupled spin-1/2 particles.
}
\label{fig_diagram}
\end{figure}

Avalanche theory has been studied mostly within  toy models, one of which was called the "Quantum Sun Model" (QSM) in~\cite{suntajs_vidmar_22, pawlik_sierant_23}, which we will mostly refer to below. 
At this moment, models like the QSM are the only examples of a many-body ergodicity breaking transition in Hamiltonian systems that have the potential to be truly well understood, as they seem well accessible both analytically and numerically.  In this paper, we want indeed to start a more detailed investigation of the QSM. The main Leitmotiv we use here, is the comparison of the QSM to a well-known model in random matrix theory, namely the "ultrametric model" (UM)~\cite{fyodorov2009anderson}.
Despite the fact that the latter model is not commonly thought of as a many-body model, we conjecture that the behaviour of the QSM and the UM is, when cast in the same language, nearly identical in the thermodynamic limit for a broad range of model parameters. 
In particular, we expect the following similarities, also sketched in Fig.~\ref{fig_diagram}.
\begin{enumerate}
    \item Both models have a natural parameter that we call $\alpha$, and that the transition occurs at $\alpha_c=1/\sqrt{2}$.
    \item On the nonergodic (localised) side of the transition, the models exhibit Fock space localization. This is characterized by the vanishing of fractal dimensions. 
    \item Both models have a continuum of critical points, that can be tuned by some natural parameters, with a natural interpretation as interpolating between the nonergodic and the ergodic side. Correspondingly, the spectral statistics at the critical points interpolate between the Poisson statistics of the nonergodic phase and the random matrix statistics of the ergodic phase.
    \item These critical points are multifractal, i.e., they are characterized by a family of critical dimensions. 
\end{enumerate}

These similarities would be in particular remarkable because the QSM, being a many-body model, has a number of disorder variables that scales like the physical volume of the system, hence like the logarithm of the Hilbert space dimension.  In contrast, the UM has a number of disorder variables that scales like the Hilbert space dimension itself. 
Yet, in a certain sense, the UM is a more natural model than the QSM to describe the coupling of a small ergodic region to a perfectly localized system; whereas the QSM artificially pretends that the coupling is only to single spins $S_i$ of the localized system, the UM describes the more realistic~\footnote{By more realistic, we mean: this is what emerges when we derive the model from the assumption that the perfectly localized region is described by many-body local integrals of motion obtained by perturbative diagonalization of the Hamiltonian} case where this particular coupling also involves all spins $S_j, j<i$, closer to the ergodic region.

In this paper, we proceed concretely as follows.
In Sec.~\ref{sec:models} we introduce the two models under consideration, the QSM and the UM.
We then introduce the indicators of the critical point in Sec.~\ref{sec:indicators}, i.e., the level spacing ratio, participation and entanglement entropies, and the Schmidt gap.
We establish two criteria of the critical point:
(a) the level spacing ratio and the entanglement entropy of the most weakly coupled spin exhibit a scale invariant point, and (b) the first derivatives of the participation and entanglement entropies exhibit a sharp peak.
In both models, these scale invariant points and the peaks of the derivatives almost perfectly coincide with the analytically predicted value for the critical point.
In Sec.~\ref{sec:multifractality} we then study Fock space localization on the nonergodic side and multifractal properties at the manifold of critical points.
To this end, we extract the fractal dimension from the scaling of participation entropies.
We argue that the degree of multifractality can be tuned by either modifying the size of the initial ergodic seed, or by the overall coupling of the ergodic seed to the remainder of the system. 
Finally, in Sec.~\ref{sec:spectrum_critical} we study properties on the critical manifold through the lens of level statistics.
We show that the values of the scale invariant level spacing ratio is close to the random matrix theory prediction if the fractal dimension is close to 1, and it decreases towards to Poisson value upon enhancing multifractality, i.e., upon decreasing the fractal dimension towards zero.
We conclude in Sec.~\ref{sec:conclusions}.

\begin{figure}[!t]
\centering
\includegraphics[width=1.00\columnwidth]{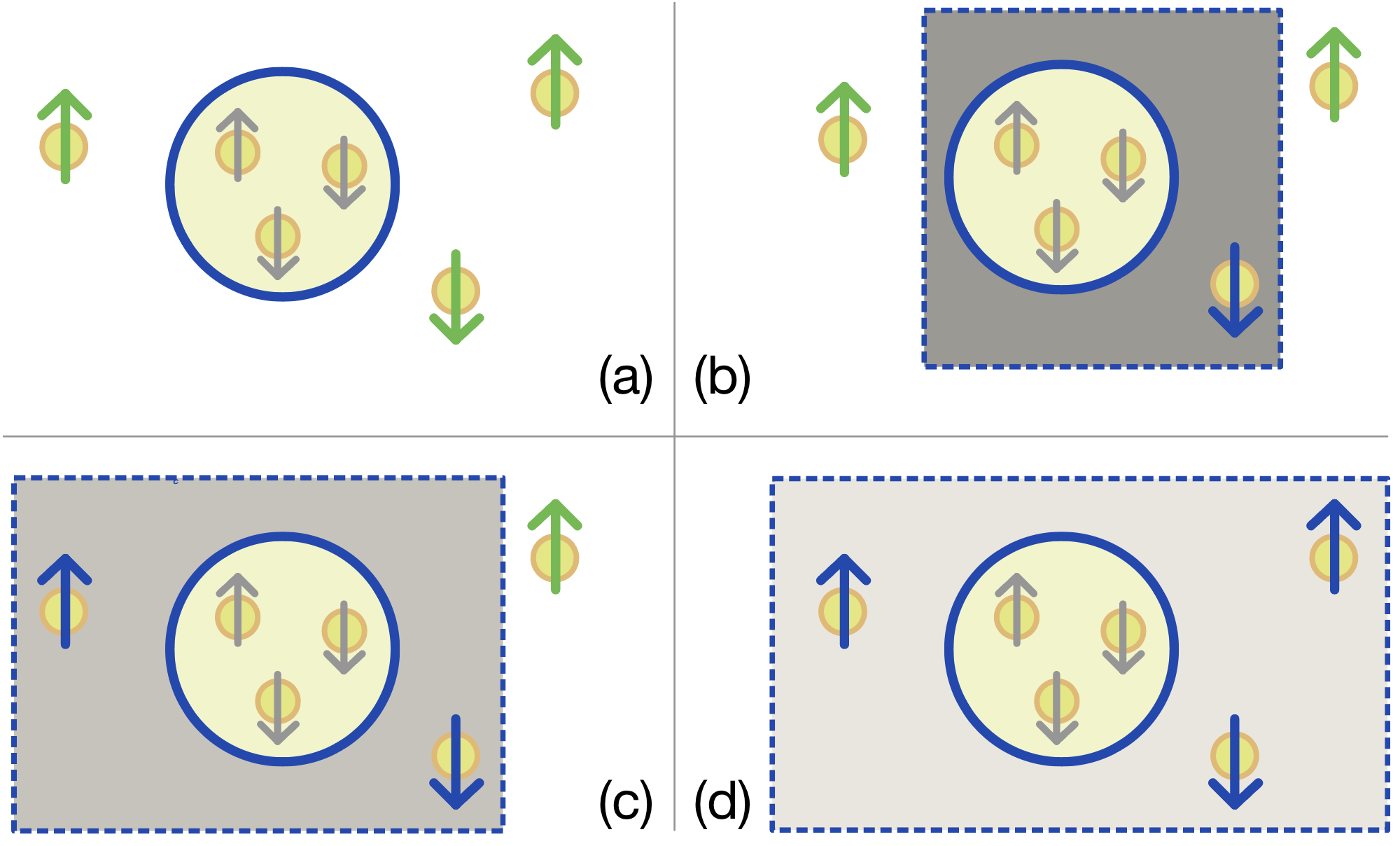}
\caption{
Sketch of the avalanche physics.
(a) The ergodic quantum dot with $N=3$ spin-1/2 particles (within a circle), which are uncoupled from the remaining $L=3$ isolated spin-1/2 particles.
The quantum avalanche is enabled by a hierarchical coupling of the dot to the remaining particles:
(b) strong coupling to the closest particle; (c) moderate coupling to the subsequent particle;
(d) weak coupling to the most distant particle.
}
\label{fig_sketch}
\end{figure}

\section{Models} \label{sec:models}

In this section, we introduce the two models studied in this work, the QSM and the UM.
In particular, we provide the definition of both models, norms of operators contained in the models, and in Appendix~\ref{sec:appendix_dos} we show their density of states.

In both models $N$ denotes the number of spin-1/2 particles within the subsystem that we denote the quantum dot.
Throughout our scaling analysis, we keep $N=\mathcal{O}(1)$ constant.
Setting $N=3$ as an example, the quantum dot is sketched as a region bounded by a blue circle in Fig.~\ref{fig_sketch}.
However, the type of interaction that gives rise to the proceeding avalanche, as sketched in Figs.~\ref{fig_sketch}(b)-\ref{fig_sketch}(d), depends on the particular model.
With the term {\it avalanche} we have in mind a process that gives rise to ergodicity of spin-1/2 particles outside the dot.
In the absence of coupling to the particles within the dot, the particles outside the dot are frozen in either of the $S_j^z = \pm 1/2$ states.
The number of particles outside the dot is denoted by $L$ and we consider the thermodynamic limit of both models by fixing $N$ and sending $L\to \infty$.

\subsection{Quantum sun model (QSM)}

The QSM Hamiltonian is given by
\begin{equation}\label{eq:def_model}
    \hat H = H_{\rm dot} + g_0 \sum\limits_{j=0}^{L-1}\alpha^{u_j}\hat{S}^x_{n_j}\hat{S}_j^x + \sum\limits_{j=0}^{L-1} h_j \hat{S}_j^z.
\end{equation}
The first term on the r.h.s.~of Eq.~(\ref{eq:def_model}), $H_{\rm dot}$, acts nontrivially on the dot degrees of freedom only.
Properties of the latter are described by a $2^N \times 2^N$ random matrix $R$
drawn from a Gaussian orthogonal ensemble (GOE), $R=\frac{1}{\sqrt{2}}\left(A + A^T\right) \in 2^N \times 2^N$, where the matrix elements $A_{i, j}$ are sampled from a normal distribution with zero mean and unit variance.
Then,
\begin{equation}\label{eq:h_dot}
    H_{\rm dot} = \frac{\gamma}{\sqrt{2^N+1}} \, R \;,
\end{equation}
such that the parameter $\gamma$ controls the bandwidth of the dot.
The prefactor $1/\sqrt{2^N + 1}$ ensures that at $\gamma=1$, $H_{\rm dot}$ has a unit Hilbert-Schmidt norm $||H_{\rm dot}||=1$, where $||H||^2 = \langle H^2\rangle = {\rm Tr}\{ H^2\}/{\cal D}$ and the Fock space dimension is ${\cal D}=2^{N+L}$ (the system has no other conservation laws apart from the total energy).

The second term on the r.h.s.~of Eq.~(\ref{eq:def_model}) describes the interaction of a strength $g_0$ between the spin-1/2 particles inside and outside the dot.
In particular, a particle $j$ outside the dot is only coupled to a randomly selected particle $n_j$ inside the dot.
The coupling strength $\alpha^{u_j}$ is tuned by the parameter $\alpha,$ and $u_j$ is the distance between a coupled particle and the dot. We sample $u_j$ from a random box distribution, $u_j \in [j - \zeta_j, j + \zeta_j]$ with $\zeta_j=0.2$ for all $j$, except for $j=0$ when $u_0 = 0$.

The third term on the r.h.s.~of Eq.~(\ref{eq:def_model}) describes the fields $h_j$ that act on spin-1/2 particles outside the dot, drawn from a random box distribution $h_j \in [W-\delta_W, W+\delta_W]$, with $W=1$ and $\delta_W=0.5.$
The matrix structure of the Hamiltonian at $N=L=3$, $\alpha=0.85$ and $g_0=\gamma=1$ is shown in Fig.~\ref{fig_matrix}(a). 

\subsection{The ultrametric model (UM)} \label{sec:models_um}

One of the main messages of this work is our conjecture that the UM of the random matrix theory (RMT) may be considered as a toy model of the avalanche theory, which retains all of the relevant physical features of the QSM while striping down all the unnecessary bits.
Several aspects of the UM have been studied in the past~\cite{fyodorov2009anderson, rushkin2011universal, vansoosten2018phase, Bogomolny2011, MendezBermudez2012, Gutkin2011, vonSoosten2017, Bogomolny2018, vonSoosten2019}, however, its connection to the avalanche theory has to our knowledge not yet been explored.

We note that the initial motivation for studying the UM was to better understand the Anderson localization transition of noninteracting particles~\cite{fyodorov2009anderson}.
Consequently, the term ''ultrametric'' corresponds to the particular geometry of a hierarchical lattice (''ultrametric lattice''), in which the metric is defined as the number of steps needed for a single particle to hop from one node to another node.
Here we are interested in many-body physics and hence we formulate the UM in the Fock space of $N+L$ spin-1/2 particles with dimension ${\cal D}=2^{N+L}$; see the definition below.
For simplicity we use the same name for the model.
One may define an ultrametric distance $d$ in our many-body model by ordering the spin-1/2 particles, cf.~Fig.~\ref{fig_sketch}, by a decreasing coupling from the ergodic quantum dot.
For simplicity, let us assume that there are no particles within the dot, i.e., $N=0$.
Then, the two spin configurations $|\sigma_1^{(a)}\sigma_2^{(a)}...\sigma_L^{(a)}\rangle$ and $|\sigma_1^{(b)}\sigma_2^{(b)}...\sigma_L^{(b)}\rangle$ are said to be at distance $d$ if $\sigma_d^{(a)} \neq \sigma_d^{(b)}$, but $\sigma_j^{(a)} = \sigma_j^{(b)}$ for all $j>d$.

The model Hamiltonian is constructed as a sum of block-diagonal random matrices $\hat{H}_k$ with $k=0, 1, \ldots, L$, where $k$ is proportional to the ultrametric distance $d$. 
At each $k,$ the matrix structure of $\hat{H}_k$ consists of $2^{L - k}$ diagonal blocks of size $2^{N + k} \times 2^{N + k}.$ In analogy with the QSM, we sample each random block independently from the GOE distribution and use normalization in accordance with Eq.~\eqref{eq:h_dot}, such that 
\begin{equation} \label{def_Hk_i}
    H^{(i)}_k = \frac{R^{(i)}}{\sqrt{2^{N + k} + 1}}, \hspace{5mm} i = 1, \ldots, 2^{L - k}\;.
\end{equation}
Here, the superscript $i$ denotes the $i$-th random block. We sample its matrix elements in analogy to the QSM, hence $R^{(i)} = \frac{1}{\sqrt{2}}(A + A^T) \in 2^{N + k} \times 2^{N+k}.$ 
The full Hamiltonian of the UM then reads
\begin{equation}\label{eq:def_rmt_model}
    \hat{H} = \hat{H}_0 + J\sum\limits_{k=1}^L \alpha^k \hat{H}_k, \hspace{5mm} \alpha\in [0, 1).
\end{equation}
The first term $\hat{H}_0$ of size $2^{N} \times 2^{N}$ models the initial quantum dot that consists of $N$ spin-1/2 particles in the absence of coupling to the external particles. The sum in the second term mimics the exponentially decaying coupling between the dot and the $k$-th localized spin-1/2 particle through the exponentially decaying values of $\alpha^k.$ Additionally, we have also included the parameter $J$ tuning the overall perturbation strength, which carries certain analogies with the parameter $g_0$ in the QSM in Eq.~(\ref{eq:def_model}).
As in the QSM, we interpret the UM as being defined in a Fock space of $N+L$ spin-1/2 particles, with ${\cal D} = 2^{N+L}$.
The matrix structure of the UM Hamiltonian at $N=L=3$, $\alpha=0.85$ and $J=1$ is shown in Fig.~\ref{fig_matrix}(b), and the density of states in both models are shown in Appendix~\ref{sec:appendix_dos}.

The UM can also be thought of as a discretized version of the power-law random banded model (PLRBM)\cite{mirlin_fyodorov_96,EversMirlin2008,Bogomolny2018}.
In the latter, the off-diagonal matrix elements $h_{i,j}$ are random numbers whose standard deviation decays with the distance $r$ from the diagonal, at large $r$, as
${\rm std}(h_{|i-j|=r}) \propto r^{-a} = 2^{-a \log_2 r}$.
The connection with the UM can be made if one approximates $\log_2 r$ with an integer, e.g., as $d= \lceil\log_2 r\rceil$, and associates $d$ with the ultrametric distance of the UM.
Then, the decay of the off-diagonal matrix elements in the PLRBM, $\propto 2^{-ad}$, should be compared with the decay in the UM at large $d$, $\propto \alpha^d/\sqrt{2^d}$, cf.~Eqs.~(\ref{def_Hk_i}) and~(\ref{eq:def_rmt_model}).
This yields the relationship between the decay exponent $a$ of the PLRBM and $\alpha$ in the UM,
\begin{equation} \label{def_a_alpha}
    a = \frac{1}{2} - \log_2 \alpha \;.
\end{equation}
It follows from Eq.~(\ref{def_a_alpha}) that the transition point $a_c=1$ in the PLRBM corresponds to the transition point $\alpha_c=1/\sqrt{2}$ in the UM.

\begin{figure}[!t]
\centering
\includegraphics[width=1.00\columnwidth]{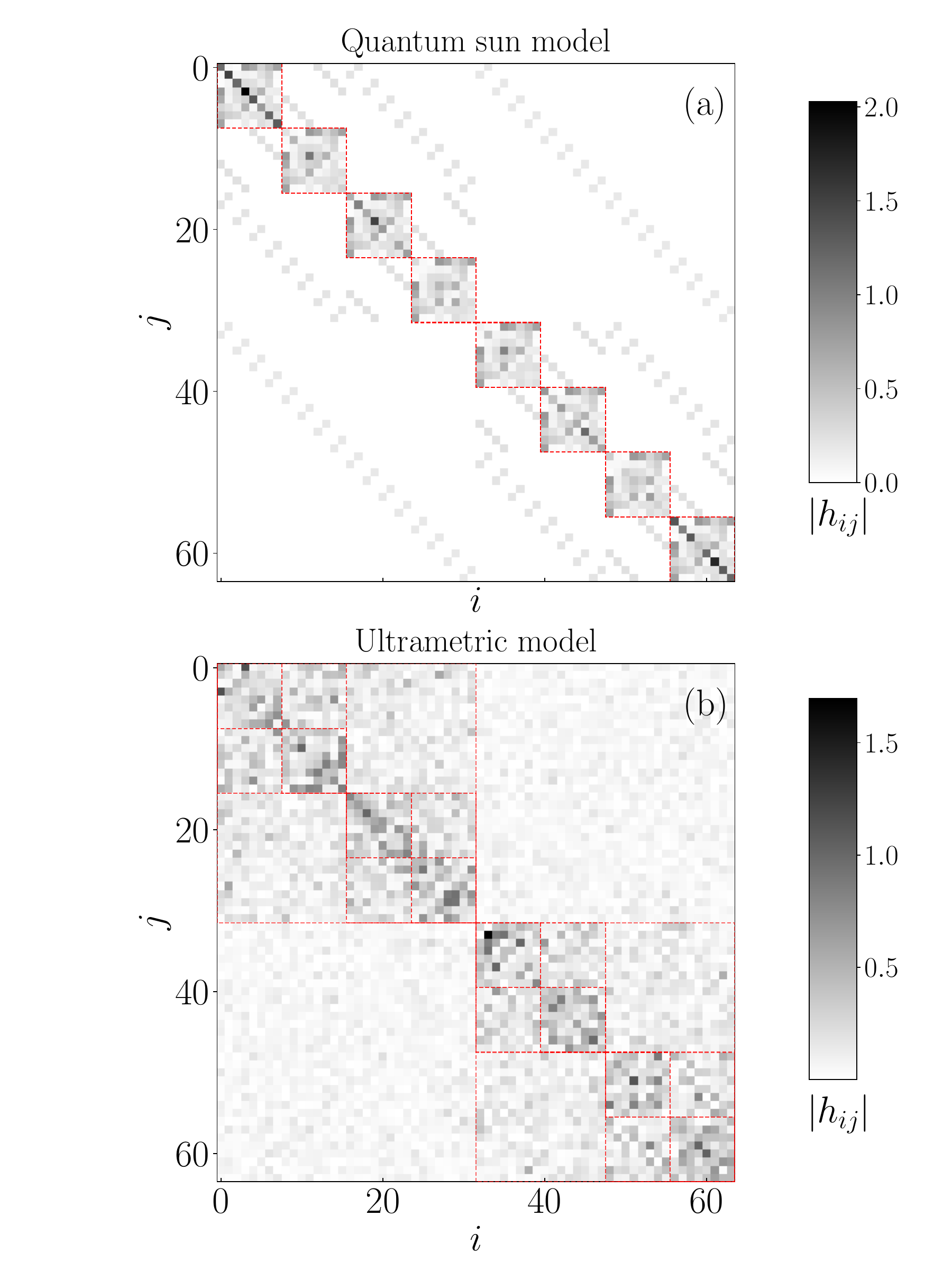}
\caption{
Hamiltonian matrix structure of (a) the quantum sun model from Eq.~(\ref{eq:def_model}), and (b) the ultrametric model from Eq.~(\ref{eq:def_rmt_model}).
In both cases, we consider $N=L=3$ and the coupling parameter $\alpha=0.85$, while the other parameters of the Hamiltonians are $g_0=\gamma=1$ in (a) and $J=1$ in (b).
A common feature of both models is the existence of $2^L = 8$ dense blocks of size $2^N \times 2^N = 8 \times 8$, denoting the ergodic quantum dot.
Outside these blocks, the matrix is sparse in (a) and dense in (b).
The matrix elements in each dense block in (a) are identical, except for their diagonal values, which are related to the distributions of random fields $h_j$, see Eq.~(\ref{eq:def_model}).
The matrix elements of each block in (b) are sampled independently.
When the size of the blocks is increased, the magnitude of their matrix elements is reduced accordingly.
}
\label{fig_matrix}
\end{figure}

\subsection{Relationship between the QSM and UM}

The matrix elements of both models, shown in Fig.~\ref{fig_matrix} at $N=L=3$, illustrate certain similarities and differences between the two models.
The matrix elements within the dot are in both cases depicted by the dense blocks of size $2^N \times 2^N = 8 \times 8$ along the diagonals.
Then, the QSM only contains two-body interactions between particles outside the dot and randomly chosen particles within the dot.
This gives rise to the overall sparse structure of the Hamiltonian matrix, shown in Fig.~\ref{fig_matrix}(a).
In the UM, on the other hand, a particle $k$ outside the dot ($k=1,2,...,L$) is effectively coupled to all particles inside the dot as well as to particles $k' < k$ outside the dot.
This gives rise to the dense Hamiltonian matrix, shown in Fig.~\ref{fig_matrix}(b).

The most important similarity of both models is that the critical point $\alpha_{\rm c}$ between an ergodic and nonergodic phase is expected to occur in the thermodynamic limit ($N$ fixed and $L\to\infty$) at the same value,
\begin{equation} \label{eq:alpha_c}
\alpha_{\rm c} = \frac{1}{\sqrt{2}} \approx 0.707 \;.
\end{equation}
This expectation is based on the hybridization condition argument~\cite{deroeck_huveneers_17}, which considers the situation when a particle $k$ outside the dot interacts via a coupling $g_k$ with the ergodic bubble consisting of $N+(k-1)$ particles.
The hybridization condition is expressed as
\begin{equation}
    {\cal G} = \frac{\langle n| \hat V |m\rangle}{\Delta} \;,
\end{equation}
where the states $|n\rangle, |m\rangle$ are tensor product states of eigenstates of the ergodic bubble and the two-level system of the spin-1/2 particle $k$.
Conjecturing that the matrix elements of eigenstates of the ergodic bubble satisfy eigenstate thermalization hypothesis~\cite{deutsch_91, srednicki_94, rigol_dunjko_08, dalessio_kafri_16}, one can approximate the matrix element as $\langle n| \hat V |m\rangle \approx g_k \sqrt{\Delta}$, where the level spacing $\Delta$ of the many-body spectrum scales as $\Delta \propto 2^{-(N+k)}$.
In both models, the coupling $g_k$ scales as $g_k \propto \alpha^k$, such that the key tuning parameter is $\alpha \in [0,1)$.
The critical point then emerges when ${\cal G} \propto \alpha^k 2^{k/2}$ remains a nonzero constant when $k\to\infty$,
which occurs at $\alpha = \alpha_{\rm c} = 1/\sqrt{2}$, as given by Eq.~(\ref{eq:alpha_c}).

We will show in this work that for system sizes amenable to state-of-the-art exact diagonalization of Hamiltonian matrices, Eq.~(\ref{eq:alpha_c}) provides an accurate estimate of the transition point for a wide range of model parameters.
In Sec.~\ref{sec:multifractality} we will also consider an example of weak coupling between the inner and outer particles in the QSM, for which the transition point in the system sizes under investigation occurs at $\alpha$ that exceeds $\alpha_{\rm c}$ from Eq.~(\ref{eq:alpha_c}).

\section{Indicators of the critical point} \label{sec:indicators}

We now turn our attention to the signatures of the critical point upon tuning the parameter $\alpha$.
We numerically study the statistical properties of the energy spectra and the properties of their corresponding energy eigenstates using full exact diagonalization of the Hamiltonian matrices from Eqs.~(\ref{eq:def_model}) and~(\ref{eq:def_rmt_model}).
In all cases under investigation, we will first focus on the UM, for which finite-size convergence of the numerical results is the most favorable, followed by the comparison with the QSM.

The goal of this section is two fold.
First, we would like to establish similarities between the two models under investigation, and show evidence that the prediction for the critical point from Eq.~(\ref{eq:alpha_c}) provides an accurate estimate for the critical point in a broad range of parameters of both models.
Then, we will compare different indicators of the critical point to explore which of them provide the most accurate prediction of the critical point, and to study for which regimes of model parameters the finite-size effects around the critical point scale most favorably.
The latter will be instrumental for the analysis of Fock space localization in the nonergodic phase and  multifractal properties at the critical point, which we will study in Sec.~\ref{sec:multifractality}.

\begin{figure*}[!t]
\centering
\includegraphics[width=1\textwidth]{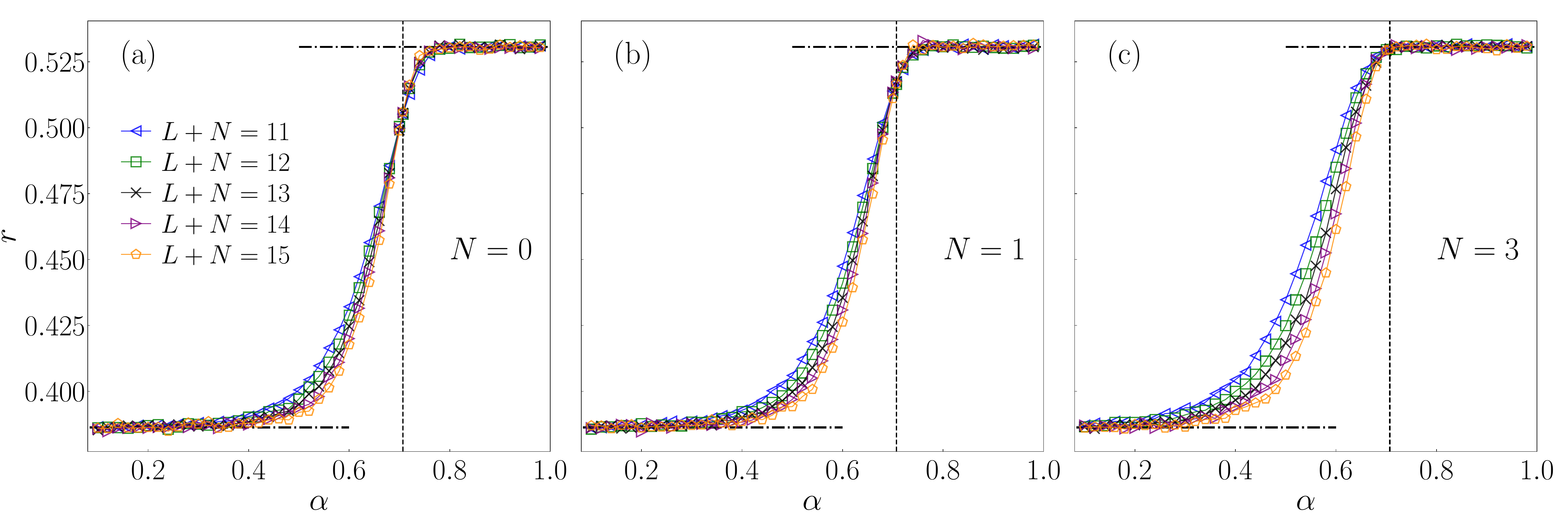}
\caption{
The average gap ratio $r$ versus $\alpha$ in the UM at $J=1$.
Panels (a), (b) and (c) show results at $N=0, \,1 $ and $3,$ respectively.
The vertical dashed line denotes the prediction $\alpha=\alpha_{\rm c}=1/\sqrt{2}$ from Eq.~(\ref{eq:alpha_c}).
%Interestingly, as $N$ is increased, the transition point exhibits a flow towards the ergodic regime, $r(\alpha_{\rm c})\to r_{\rm GOE}.$ 
The GOE and Poisson limits $r_{\rm GOE}$ and $r_{\rm Poisson}$ are denoted by the upper and lower horizontal dashed-dotted lines, respectively.
Results for $r$ are averaged over $N_\mathrm{samples}=1000,\, 400, \,300$ Hamiltonian realizations for $L + N\leq 13\,, L + N=14, \, L + N=15, $ respectively. 
}
\label{fig_r1_toy}
\end{figure*}

\begin{figure}[!t]
\centering
\includegraphics[width=0.48\textwidth]{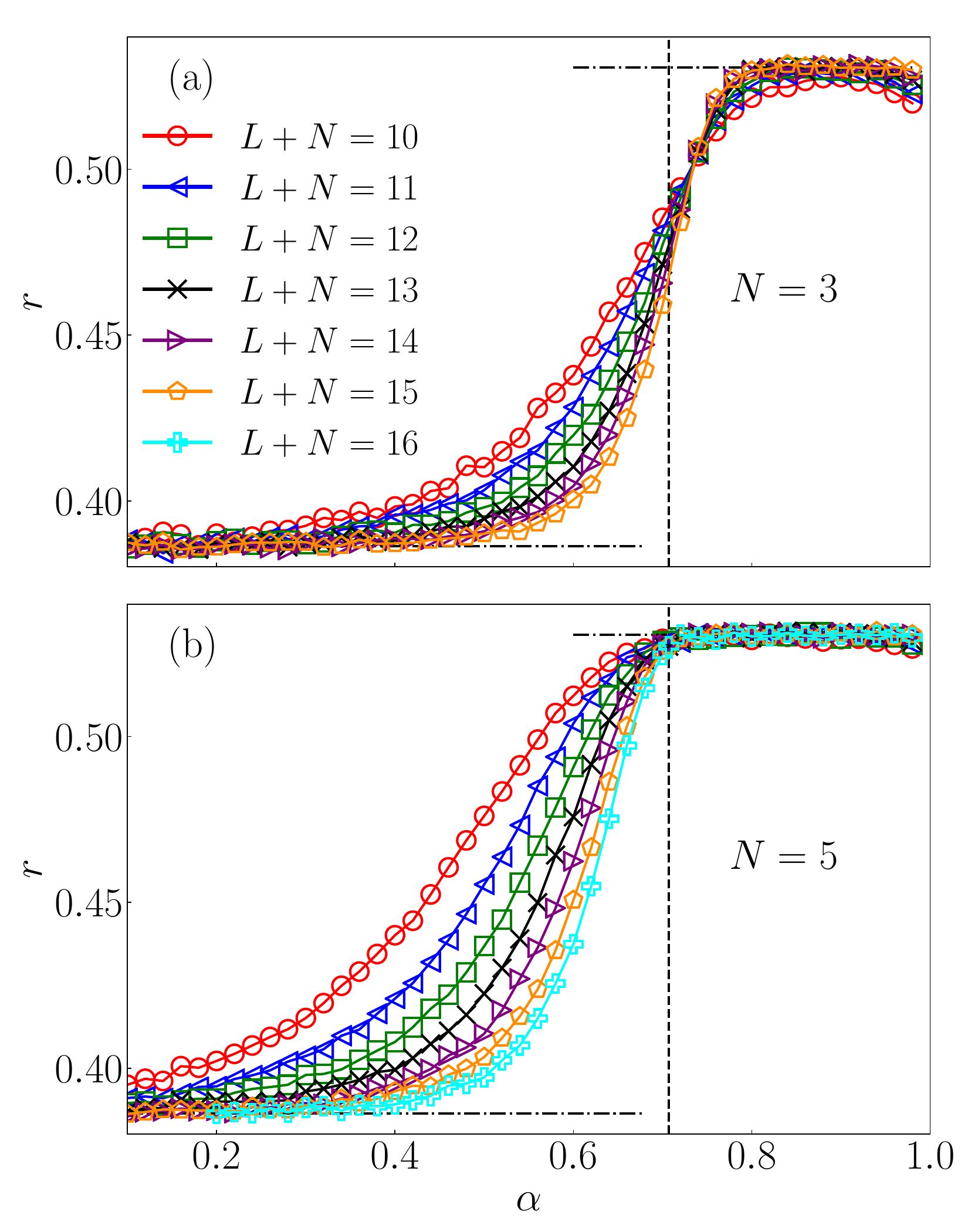}
\caption{
The average gap ratio $r$ versus $\alpha$ in the QSM at $g_0=\gamma=1$.
(a) $N=3$ and (b) $N=5$.
The vertical dashed line denotes the prediction $\alpha=\alpha_{\rm c}=1/\sqrt{2}$ from Eq.~(\ref{eq:alpha_c}).
The GOE and Poisson limits $r_{\rm GOE}$ and $r_{\rm Poisson}$ are denoted by the upper and lower horizontal dashed-dotted lines, respectively.
Results for $r$ are averaged over $N_\mathrm{samples}=500$ Hamiltonian realizations.
}
\label{fig_r1_sun}
\end{figure}

\subsection{The average ratio of the adjacent level spacings} \label{sec:gap}

We first proceed by analysing the average ratio of the adjacent level spacings $r$~\cite{oganesyan_huse_07}, shortly the average gap ratio.
For a target eigenlevel $n$, the ratio $\tilde{r}_n$ is defined as
\begin{equation}\label{eq:defr}
    \tilde{r}_n = \frac{\min\{\delta E_n, \delta E_{n-1}\}}{\max\{\delta E_n, \delta E_{n-1}\}} = \min\{r_n, r_n^{-1}\}\;,
\end{equation}
where $\delta E_n = E_{n+1} - E_n$ is the level spacing between the eigenlevels $n+1$ and $n$, respectively, while $r_n$ is the ratio of the consecutive level spacings, $r_n = \delta E_n/\delta E_{n-1}.$ By definition, $\{\tilde{r}_n\}$ only assume values from the interval $[0, 1],$ and hence no unfolding procedure is needed to eliminate the influence of finite-size effects through the local density of states. To obtain the average value $r$,
\begin{equation} \label{eq:def_r}
    r = \langle \langle \tilde{r}_n \rangle_n \rangle_H \;,
\end{equation}
we first average over $N_{\rm eig}=500$ eigenstates near the center of the spectrum for each Hamiltonian realization, denoted by $\langle \cdots \rangle_n$ in Eq.~(\ref{eq:def_r}), and then over an ensemble of spectra for different Hamiltonian realizations, denoted by $\langle \cdots \rangle_H$. In the ergodic regime, $r$ assumes the GOE value $r_{\rm GOE}\approx 0.5307$~\cite{atas_bogomolny_13}, while the prediction for energy levels with Poissonian statistics is $r_{\rm Poisson} = 2\ln 2 - 1 \approx 0.3863$~\cite{oganesyan_huse_07}. 

The results for $r$ versus $\alpha$ are shown for the UM (at $J=1$) in Fig.~\ref{fig_r1_toy} and for the QSM (at $g_0=\gamma=1$) in Fig.~\ref{fig_r1_sun}.
There are two main messages from these results.

The first result is that the critical point is rather accurately predicted for both models by $\alpha_{\rm c} = 1/\sqrt{2}$ from Eq.~(\ref{eq:alpha_c}).
The location of the critical point is estimated by a crossing point or $r$ versus $\alpha$ at different values of $L$, which signals scale invariance of $r$ at the critical point.
We will show in the next sections that this approach to detect the critical point agrees very well with those based on the analysis of participation and entanglement entropies.

The second result is that by increasing $N$ the spectral statistics at the critical point gets closer to the GOE value $r_{\rm GOE}$. This is most prominent at $N=3$ in the UM, see Fig.~\ref{fig_r1_toy}(c), and at $N=5$ in the QSM, see Fig.~\ref{fig_r1_sun}(b), for which the point $\alpha=\alpha_{\rm c}$ signals the breakdown of $r$ from $r_{\rm GOE}$.
In these cases, therefore, the criterion for the critical point based on the breakdown of $r$ from $r_{\rm GOE}$ replaces the criterion based on the emergence of a scale invariant crossing point of the $r$ values.

Figure~\ref{fig_r1_sun} also reveals an important numerical detail about the QSM.
Namely, at $N=3$ the crossing point of the $r$ values emerges at $\alpha$ that is slightly larger than $\alpha_{\rm c}$, see Fig.~\ref{fig_r1_sun}(a), while at $N=5$ the breakdown point of $r$ from $r_{\rm GOE}$ is accurately predicted by $\alpha_{\rm c}$.
This property was already noticed in Ref.~\cite{suntajs_vidmar_22} that considered $N=3$ and in Ref.~\cite{hopjan2023} that considered $N=5$.
We conclude from these results that by increasing $N$ in the QSM, the transition point for the available system sizes gets more accurately predicted by $\alpha_{\rm c}$ from Eq.~(\ref{eq:alpha_c}).
Still, since the thermodynamic limit is obtained by increasing $L\to\infty$ and hence $L$ is expected to be larger than $N$, the value of $N$ in the actual calculations should not be too large.
In what follows, we set $N=5$ as a compromise to obtain optimal convergence of finite-size effects in the QSM.

In Figs.~\ref{fig_r2_toy} and~\ref{fig_r2_sun} of Sec.~\ref{sec:multifractality} we will further show that decreasing $J$ and $g_0$ in the UM and the QSM, respectively, further shifts the $r$ values at the critical point closer to the Poisson value $r_{\rm Poisson}$.
This insight will be used to tune the multifractal properties of the eigenstates at the critical point.

\begin{figure*}[!t]
\centering
\includegraphics[width=0.850\textwidth]{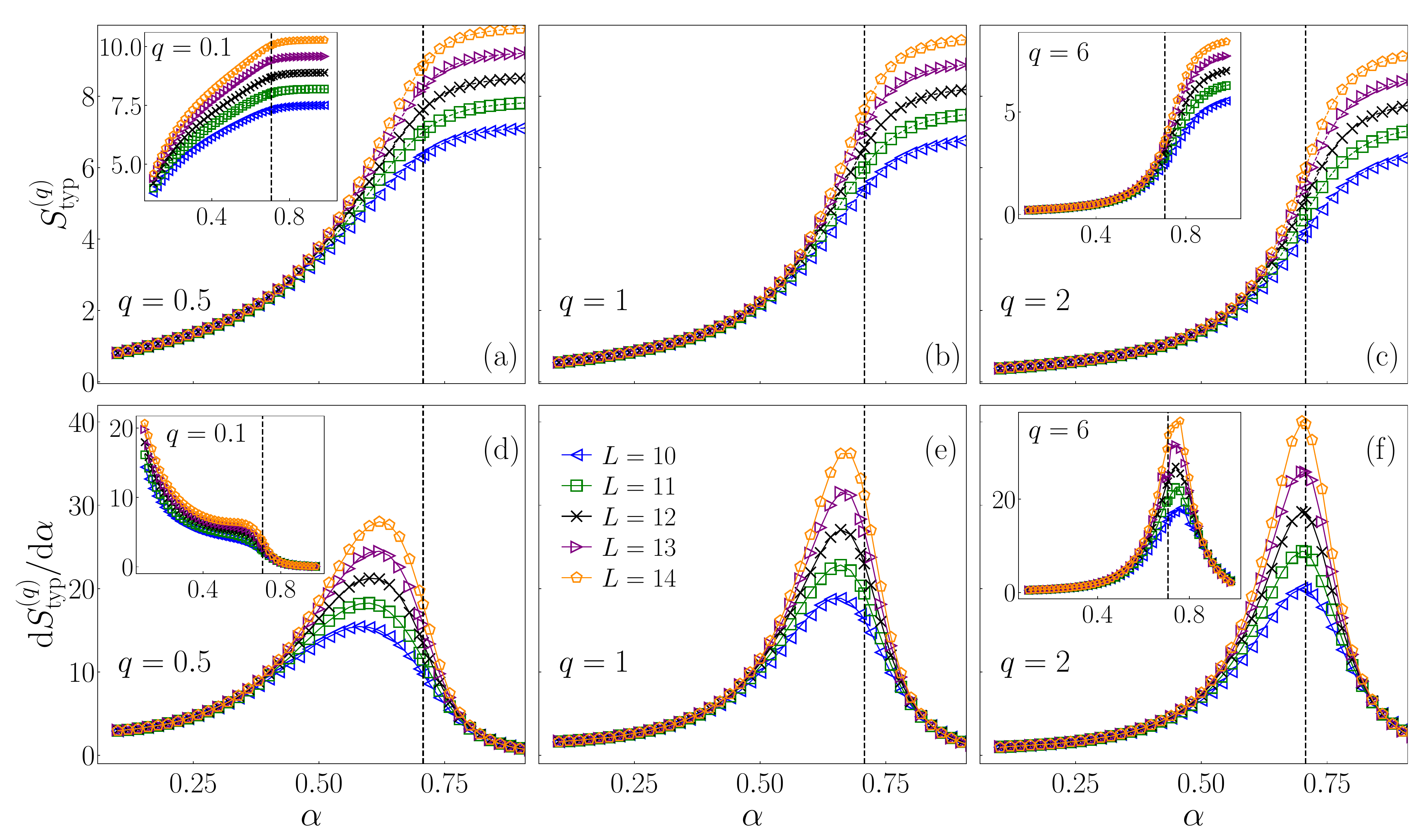}
\caption{
Participation entropies $S_{\rm typ}^{(q)}$ [panels (a-c)] and their derivatives ${\rm d} S_{\rm typ}^{(q)}/{\rm d}\alpha$ [panels (d-f)] in the UM at $J=1$ and $N=1$.
(a) and (d): $q=0.5$ in the main panel and $q=0.1$ in the insets.
(b) and (e): $q=1$.
(c) and (f): $q=2$ in the main panel and $q=6$ in the inset.
Vertical dashed lines denote $\alpha=\alpha_{\rm c}$ from Eq.~(\ref{eq:alpha_c}).
Results for $S_{\rm typ}^{(q)}$ are averaged over $N_\mathrm{samples}=1000,\, 400, \,300$ Hamiltonian realizations for $L\leq 12\,, L=13, \, L=14, $ respectively.
Averaging over the eigenstates, $\langle \cdots \rangle_n,$ was performed over $N_{\mathrm{eig}} = 1000$ eigenstates near the center of the spectrum for each data point.
}
\label{fig_sp2_toy}
\end{figure*}

\begin{figure*}[!t]
\centering
\includegraphics[width=0.850\textwidth]{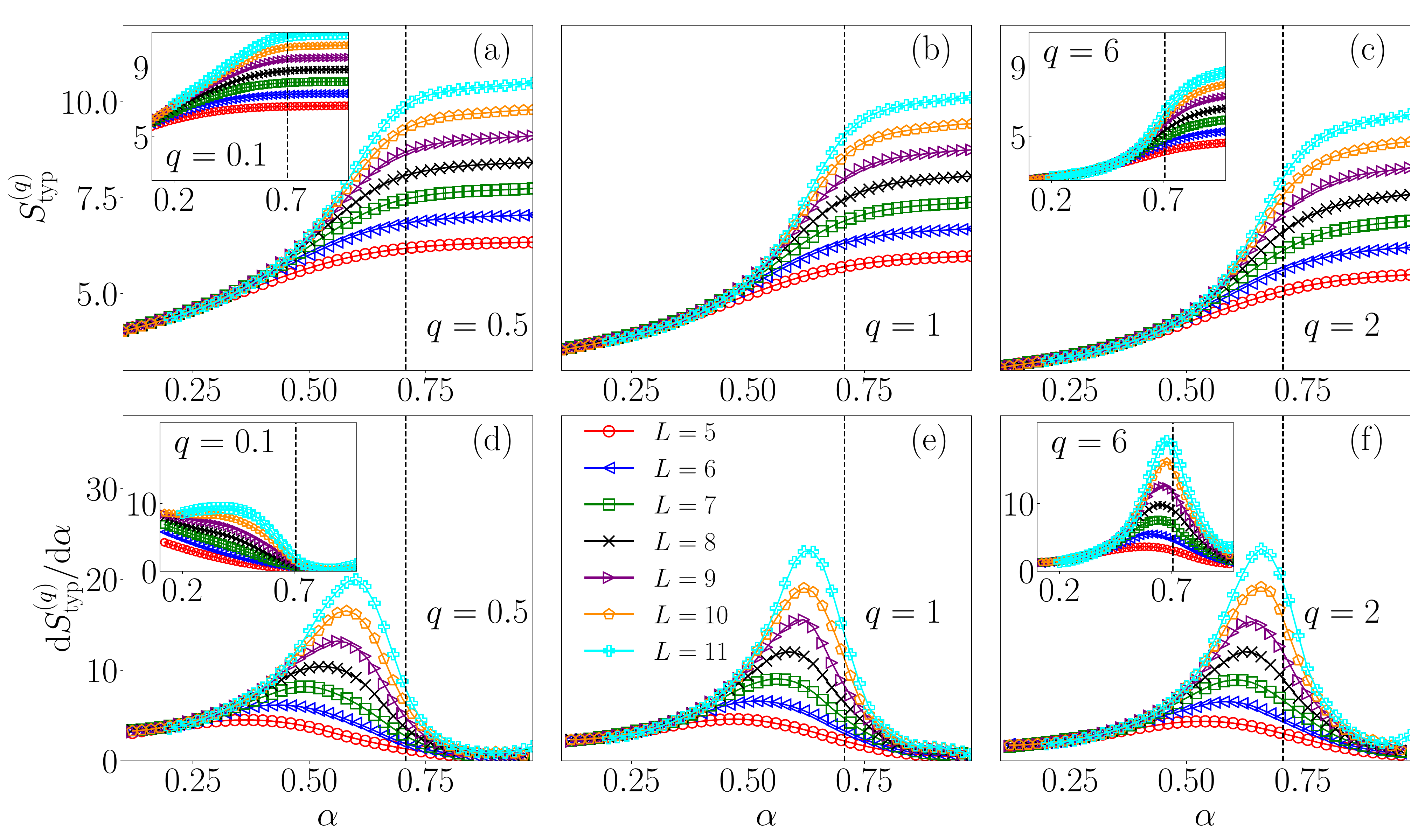}
\caption{
Participation entropies $S_{\rm typ}^{(q)}$ [panels (a-c)] and their derivatives ${\rm d}S_{\rm typ}^{(q)}/{\rm d}\alpha$ [panels (d-f)] in the QSM at $g_0=\gamma=1$ and $N=5$.
(a) and (d): $q=0.5$ in the main panel and $q=0.1$ in the insets.
(b) and (e): $q=1$.
(c) and (f): $q=2$ in the main panel and $q=6$ in the inset.
Vertical dashed lines denote $\alpha=\alpha_{\rm c}$ from Eq.~(\ref{eq:alpha_c}).
Results for $S_{\rm typ}^{(q)}$ are averaged over $N_\mathrm{samples} = 500$ Hamiltonian realizations.
Averaging over the eigenstates, $\langle \cdots \rangle_n,$ was performed over $N_{\mathrm{eig}} = 500$ eigenstates near the center of the spectrum for each data point.
}
\label{fig_sp2_sun}
\end{figure*}

\subsection{The inverse participation ratio and the participation entropy} \label{sec:ipr}

We next focus our attention towards the properties of the Hamiltonian eigenstate wavefunctions.
To that end, we calculate the inverse participation ratio (IPR) for a selected number of Hamiltonian eigenstates $\ket{n}.$
The IPR, generalized to an arbitrary index $q$, is defined as
\begin{equation}\label{eq:def_IPR}
    P_q^{-1}(\ket{n}) = \sum_{i=1}^{\mc D}|\braket{i|n}|^{2q}\;.
\end{equation}
The generalized IPR as such is basis dependent. Here, $\ket{i}$ are the states in the computational basis, i.e, the basis in which $\hat S_i^z$ operators are diagonal, $\ket{n}$ is a Hamiltonian eigenstate at the corresponding energy $E_n$, and $q$ is a parameter of the calculation. 

For convenience, the quantity that we actually study is the participation entropy, which is defined as the logarithm of the IPR. Here we mainly focus on the participation entropy of the typical IPR,
\begin{equation}\label{eq:def_participation_entro}
    S_{\rm typ}^{(q)} = \frac{1}{1-q}\langle \langle \ln P_q^{-1} \rangle_n\rangle_{H}\;,
\end{equation}
where the brackets $\langle\cdots\rangle_n$ and $\langle\cdots\rangle_H$ denote the averages over Hamiltonian eigenstates at a fixed realization and different Hamiltonian realizations, respectively. 
The limiting value of $S_{\rm typ}^{(q)}$ as $q \to 1$ is the von Neumann participation entropy, 
\begin{equation} \label{eq:def_S1typ}
S_{\rm typ}^{(1)} = - \Big\langle\Big\langle \sum_{i=1}^{\mc D}|\braket{i|n}|^{2} \ln  |\braket{i|n}|^{2} \Big\rangle_n \Big\rangle_H\;.
\end{equation}
We note that we have also studied the participation entropy of the average IPR, $S_{}^{(q)} = \frac{1}{1-q}\ln \langle \langle P_q^{-1} \rangle_n\rangle_{H}$, with no qualitative physical differences (not shown here). 

The goal of this section is to explore to which extent one can use the IPR-based quantities to pinpoint the critical point.
Figures~\ref{fig_sp2_toy}(a)-\ref{fig_sp2_toy}(c), and the corresponding insets, show $S_{\rm typ}^{(q)}$ versus $\alpha$ at various $q$ in the UM at $J=1$ and $N=1$.
They behave according to the expectations: $S_{\rm typ}^{(q)}$ approaches in the limit $\alpha\to 0$ an $L$-independent constant, and it increases (roughly linearly) with $L$ in the ergodic phase at $\alpha > \alpha_{\rm c}$.
However, based on the analysis of $S_{\rm typ}^{(q)}$ only, it appears to be impossible to determine the critical point without any {\it a priori} knowledge of it.

In Figs.~\ref{fig_sp2_toy}(d)-\ref{fig_sp2_toy}(f) we show that a valuable information about the critical point can be obtained by calculating the derivative of the participation entropy, ${\rm d}S_{\rm typ}^{(q)}/{\rm d}\alpha$, and plot it versus $\alpha$.
This analysis is inspired by the recent results for the Anderson models~\cite{suntajs_prosen_21, suntajs_prosen_23}, which showed that the peak of ${\rm d}S_{\rm typ}^{(q)}/{\rm d}W$ (where $W$ denotes the disorder amplitude) is located very close to the Anderson localization transition in three dimensions~\cite{suntajs_prosen_21}, and it drifts to zero in two dimensions~\cite{suntajs_prosen_23}.
A similar tendency is also observed in the UM at $J=1$ and $N=1$, shown in Figs.~\ref{fig_sp2_toy}(d)-\ref{fig_sp2_toy}(f).
Namely, the peak in ${\rm d}S_{\rm typ}^{(q)}/{\rm d}\alpha$ is, for most values of $q$, located very close to the predicted critical point $\alpha=\alpha_{\rm c}$ from Eq.~(\ref{eq:alpha_c}). 

As a technical remark, we calculate the derivatives ${\rm d}S_{\rm typ}^{(q)}/{\rm d}\alpha$ by first interpolating the raw data for $S_{\rm typ}^{(q)}$ with cubic splines using the \url{UnivariateSpline} function from the \url{scipy.interpolate} library. Then, using the tools available within the same interpolating function, we also evaluate the first derivatives. We use the same procedure also to evaluate the derivatives of the Rényi entanglement entropies in Sec.~\ref{sec:Renyi}.

The $q$-dependence of the results in Figs.~\ref{fig_sp2_toy}(d)-\ref{fig_sp2_toy}(f) reveals rich information about the behavior of participation entropies close to the critical point.
We observe that at $q=2$, see Fig.~\ref{fig_sp2_toy}(f), the peak location coincides with $\alpha=\alpha_{\rm c}$ to extremely high accuracy even in small systems.
This remarkable property is also observed at higher $q$, see the inset of Fig.~\ref{fig_sp2_toy}(f), while at smaller $q$ this high precision is lost, see the results in Figs.~\ref{fig_sp2_toy}(d)-\ref{fig_sp2_toy}(e).
In the localized regime of the PLRBM (i.e., at $a>1$), it was shown that the generalized IPR at $q<1/2$ may not exhibit localization at arbitrary $a>1$ as a consequence of power-law localization~\cite{mirlin_fyodorov_96,EversMirlin2008}.
Based on the relationship between the PLRBM and UM discussed in Sec.~\ref{sec:models_um}, we expect similar features to emerge in the UM as well.
This is consistent with the observation in the inset of Fig.~\ref{fig_sp2_toy}(d), which suggests that at $q\ll 1$ in the UM it becomes nearly impossible to detect the critical point since ${\rm d}S_{\rm typ}^{(q)}/{\rm d}\alpha$ exhibits a shoulder in the vicinity of the critical point rather than a sharp peak.

While the results in Fig.~\ref{fig_sp2_toy} were obtained for the UM at $N=1$, in Fig.~\ref{fig_sp1_toy} of Appendix~\ref{sec:appendix_Sp} we also show the results (at $q=2$) for other values of $N$.
We observe that the finite-size effects scale most favorably at $N=1$, i.e., at this value of $N$ the peak of ${\rm d}S_{\rm typ}^{(q)}/{\rm d}\alpha$ emerges almost exactly at $\alpha=\alpha_{\rm c}$ already in small systems of $N+L=11$ spin-1/2 particles.
We hence consider the UM at $N=1$ as the model that is best suited for the analysis of the critical behavior in finite systems, and we only study the $N=1$ case further on.

In Fig.~\ref{fig_sp2_sun} we complement the results for the participation entropies in the UM by showing the results for the QSM at $g_0=\gamma=1$ and $N=5$.
Qualitatively, the results in Fig.~\ref{fig_sp2_toy} and~\ref{fig_sp2_sun} are rather similar, which is the main message of these analyses.
Still, the accuracy to determine the location of the critical point via the peak of ${\rm d}S_{\rm typ}^{(q)}/{\rm d}\alpha$ is in the QSM not as high as in the UM.
Figure~\ref{fig_sp2_sun}(f) shows that again the most accurate results are obtained in the large $q$ regime, $q \gtrsim 2$.
In case the precise position of the critical point is not known in advance, a possible method to improve the prediction for the critical point (that we do not pursue here) is to extract the position of the peak for each $L$, and then scale these values to the limit $L\to\infty$.

To summarize the results for the participation entropies, we note that its derivative w.r.t.~the tuning parameter of the transition, ${\rm d}S_{\rm typ}^{(q)}/{\rm d}\alpha$, exhibits a peak that we expect is located at the critical point in the thermodynamic limit.
An additional insight that we obtained from our analysis is that in finite systems amenable to exact diagonalization, the most accurate prediction for the critical point are obtained at large values of $q \gtrsim 2$.
On the other hand, the results in the opposite limit $q \ll 1$ are, at least for the available system sizes, not very useful for determining the critical point.

\begin{figure*}[!t]
\centering
\includegraphics[width=0.85\textwidth]{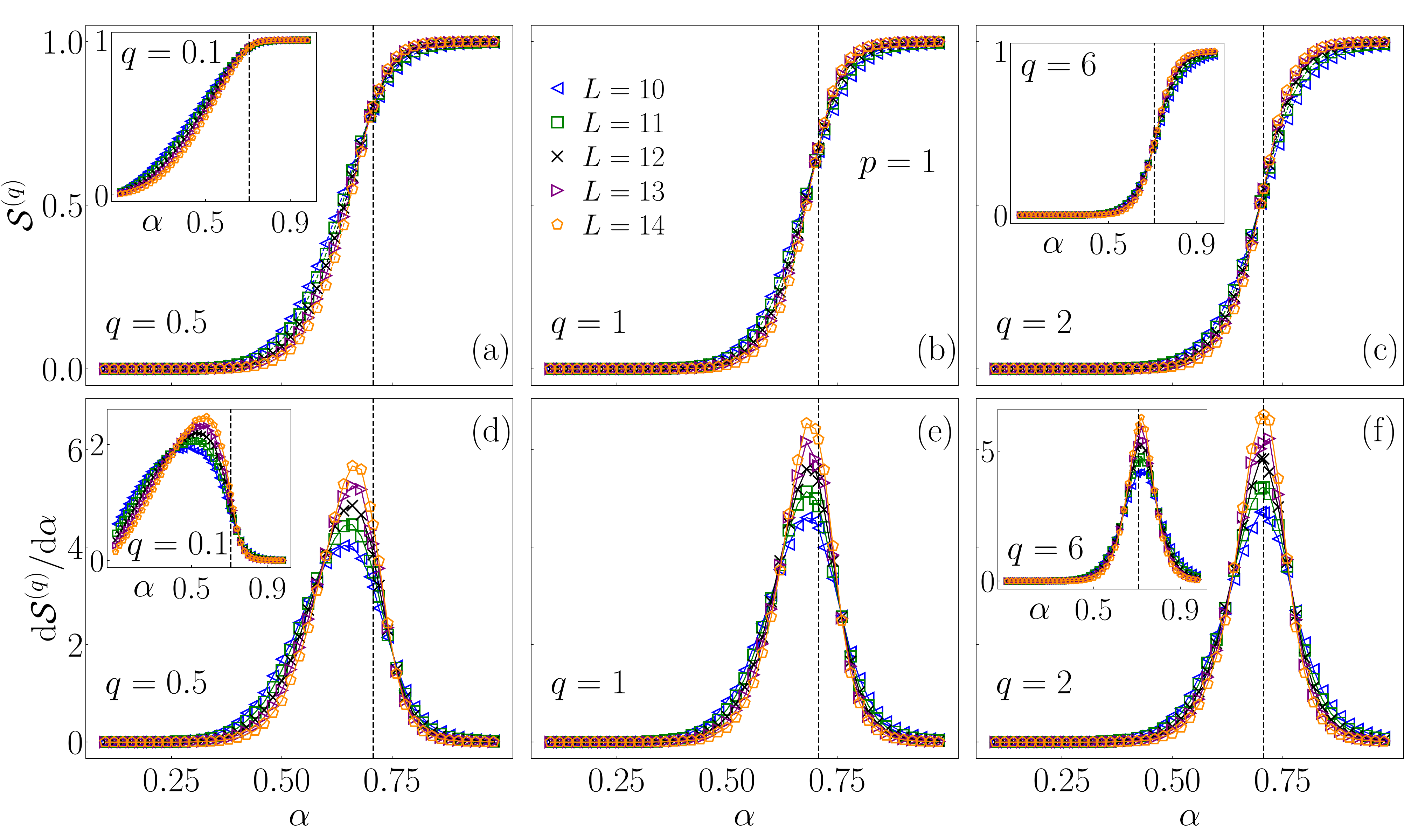}
\caption{
The Rényi entanglement entropies ${\cal S}^{(q)}$ [panels (a-c)] and their derivatives ${\rm d}{\cal S}^{(q)}/{\rm d}\alpha$ [panels (d-f)] in the UM at $J=1$, $N=1$ and $p=1$.
(a) and (d): $q=0.5$ in the main panel and $q=0.1$ in the insets.
(b) and (e): $q=1$.
(c) and (f): $q=2$ in the main panel and $q=6$ in the inset.
Vertical dashed lines denote $\alpha=\alpha_{\rm c}$ from Eq.~(\ref{eq:alpha_c}).
Results for ${\cal S}^{(q)}$ are averaged over $N_\mathrm{samples}=1000,\, 400, \,300$ Hamiltonian realizations for $L\leq 12\,, L=13, \, L=14, $ respectively. 
Averaging over the eigenstates, $\langle \cdots \rangle_n,$ was performed over $N_{\mathrm{eig}} = 1000$ eigenstates near the center of the spectrum for each data point.
}
\label{fig_sr1_toy}
\end{figure*}

\begin{figure*}[!t]
\centering
\includegraphics[width=0.85 \textwidth]{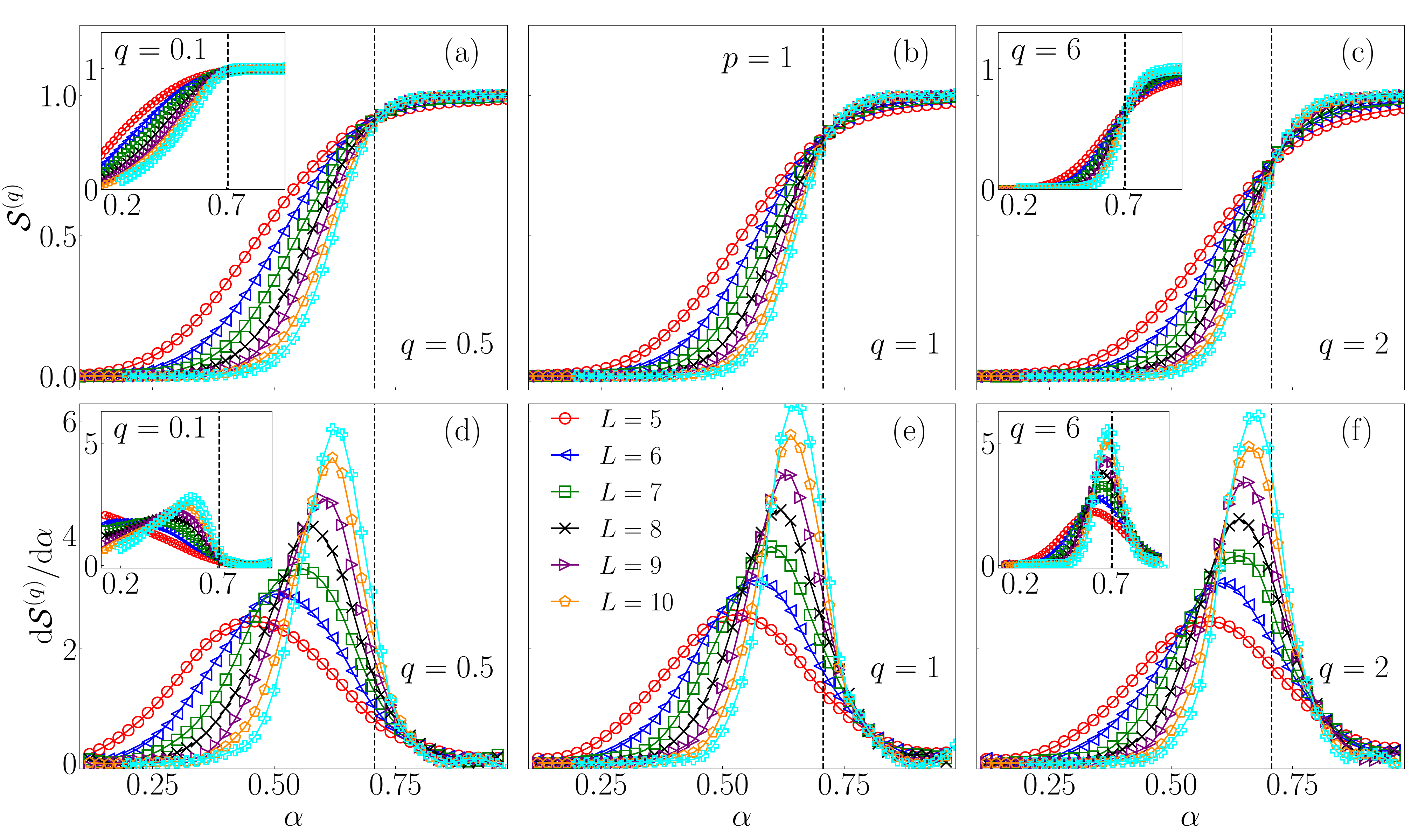}
\caption{
The Rényi entanglement entropies ${\cal S}^{(q)}$ [panels (a-c)] and their derivatives ${\rm d}{\cal S}^{(q)}/{\rm d}\alpha$ [panels (d-f)] in the QSM at $g_0=\gamma=1$, $N=5$ and $p=1$.
(a) and (d): $q=0.5$ in the main panel and $q=0.1$ in the insets.
(b) and (e): $q=1$.
(c) and (f): $q=2$ in the main panel and $q=6$ in the inset.
Vertical dashed lines denote $\alpha=\alpha_{\rm c}$ from Eq.~(\ref{eq:alpha_c}).
Results for ${\cal S}^{(q)}$ are averaged over $N_\mathrm{samples}=500$ Hamiltonian realizations.
Averaging over the eigenstates, $\langle \cdots \rangle_n,$ was performed over $N_{\mathrm{eig}} = 500$ eigenstates near the center of the spectrum for each data point.
}
\label{fig_sr1_sun}
\end{figure*}

\subsection{Rényi entanglement entropies of eigenstates} \label{sec:Renyi}

While the analysis of the participation entropies in the previous section turned out to be quite useful for the determination of the critical point, we here complement these results by studying the entanglement based measures.
The participation entropies contain information about the whole wavefunction, and hence they may not be very sensitive to the properties of the particles that are ''most distant'' from the ergodic quantum dot.
(With ''most distant'' we have in mind the particles that are most weakly coupled to the ergodic quantum dot.)
On the other hand, the entanglement entropies allow for selecting arbitrary subsystems.
Here we focus on the entanglement properties of most distant spin-1/2 particles.
A physical motivation for this choice is that the ergodicity breaking phase transition within the avalanche theory is expected to occur when the avalanche fails to thermalize the most distant particles, and hence the change of their properties contains crucial information about the transition.

We consider the eigenstate entanglement entropies of subsystems that consist of $p$ most distant spin-1/2 particles.
The Fock space $\mc H$ of the studied Hamiltonian carries a tensor product structure
\begin{equation}\label{eq:tensor_product_h}
    \mathcal{H} = \mathcal{H}_N \otimes \mathcal{H}_1 \otimes \ldots \otimes \mathcal{H}_L,
\end{equation}
where $\mathcal{H}_N$ refers to the dot degrees of freedom and $\mathcal{H}_i$ to the $i$-th spin-1/2 particle outside the dot.
We partition the system into two partitions $A_p$ and $B_p$, such that $\mathcal{H} = \mathcal{H}_{A_p} \otimes \mathcal{H}_{B_p}.$ Here, we take $B_p$ to be a collection of $p$ spins farthermost from the dot,
$$
B_p = \{L - p + 1, \ldots L \},
$$
while $A_p$ denotes the remainder of the system that consists of $N+L-p$ particles.
In the following, unless necessary, we shall omit the subscript $p$ when referring to the partitions. 

For the density matrix $\hat{\rho}$ of the full system, we obtain the reduced density matrix of the subsystem $B$ by tracing
out the degrees of freedom in $A$,
\begin{equation}\label{eq:def_rdm}
    \hat{\rho}_B = {\rm Tr}_A{\hat{\rho}}, \hspace{5mm} {\rm Tr} \hat{\rho}_B = 1 \;,
\end{equation}
where $\hat \rho = |n\rangle\langle n |$ is the density matrix associated to the Hamiltonian eigenstate $|n\rangle$.
Upon diagonalization of $\hat{\rho}_B$ we obtain the eigenvalue spectrum of the reduced density matrix, which we denote by $\{\lambda_1, \ldots, \lambda_i, \ldots,\lambda_{{\mc D}_B}\}$, where $\lambda_i \geq \lambda_{i+1}.$ Here, ${\mc D}_B$ is the reduced Fock space dimension which equals $2^p$ in our analysis.
The $q$-th Rényi eigenstate entanglement entropy, which is normalized such that its maximum value is bounded from above by 1, is then defined as
\begin{equation}\label{eq:def_renyi}
    {\mathcal S}^{(q)} = \frac{1}{\ln{\mathcal{D}_B}} \frac{1}{1 - q} \Big\langle \Big\langle \ln \sum\limits_{i=1}^{\mathcal{D}_B} \lambda_i^q \Big\rangle_n\Big\rangle_{H}\;, 
\end{equation}
with $q > 0$ and $q\neq 1.$
The limiting value of ${\mathcal S}^{(q)}$ as $q \to 1$ is the von Neumann eigenstate entanglement entropy, 
\begin{equation}
    {\cal S}^{(1)} = - \frac{1}{\ln{\mathcal{D}_B}} \Big\langle \Big\langle \sum\limits_{i=1}^{\mathcal{D}_B} \lambda_i \ln\lambda_i \Big\rangle_n\Big\rangle_{H}\;.
\end{equation}
As in Eqs.~(\ref{eq:def_r}),~(\ref{eq:def_participation_entro}) and~(\ref{eq:def_S1typ}), the averaging is performed over Hamiltonian eigenstates and different Hamiltonian realizations.
At $p=1$ one can get further analytical insight into the entanglement entropies, which we present in Appendix~\ref{sec:appendix_Sr}.

Results for the eigenstate entanglement entropies ${\mathcal S}^{(q)}$ at $p=1$ are shown for the UM in Fig.~\ref{fig_sr1_toy}.
They are consistent with the limiting behaviors described in Eqs.~(\ref{eq:Sq_small}) and~(\ref{eq:Sq_large}) of Appendix~\ref{sec:appendix_Sr}, i.e., ${\mathcal S}^{(q\to 0)}$ is very close to 1 at $\alpha=\alpha_{\rm c}$, see the inset of Fig.~\ref{fig_sr1_toy}(a), and ${\mathcal S}^{(q\to \infty)}$ at $\alpha=\alpha_{\rm c}$ is considerably smaller than 1, see the inset of Fig.~\ref{fig_sr1_toy}(c).

The eigenstate entanglement entropies ${\mathcal S}^{(q)}$ exhibit an important advantage when compared to the participation entropies $S_{\rm typ}^{(q)}$, namely, the critical point can readily be estimated to high accuracy from the crossing point of ${\mathcal S}^{(q)}$ versus $\alpha$ at different $L$, see Figs.~\ref{fig_sr1_toy}(a)-\ref{fig_sr1_toy}(c).
Moreover, its derivative, ${\rm d}{\mathcal S}^{(q)}/{\rm d}\alpha$, also exhibits a peak that is located very close to the predicted critical point $\alpha=\alpha_{\rm c}$.
We again observe that the location of the peak almost exactly coincides with $\alpha_{\rm c}$ at large $q$, see Fig.~\ref{fig_sr1_toy}(f), while at small $q$ the agreement is less accurate, see Fig.~\ref{fig_sr1_toy}(d).
To evaluate the derivatives of the entanglement entropies, we use the same procedure as the one outlined in Sec.~\ref{sec:ipr}.

Analogous results for the eigenstate entanglement entropies ${\mathcal S}^{(q)}$ at $p=1$ are shown for the QSM in Fig.~\ref{fig_sr1_sun}.
Also in this case, there exist a scale invariant (i.e., $L$ independent) point of ${\mathcal S}^{(q)}$ that is almost exactly located at the critical point $\alpha=\alpha_{\rm c}$, see Figs.~\ref{fig_sr1_sun}(a)-\ref{fig_sr1_sun}(c).
The derivative ${\rm d}{\mathcal S}^{(q)}/{\rm d}\alpha$ exhibits a peak very close to the critical point, see Figs.~\ref{fig_sr1_sun}(d)-\ref{fig_sr1_sun}(f).
Nevertheless, the location of the peak in the QSM does not coincide with $\alpha=\alpha_{\rm c}$ as accurately as in the UM shown in Figs.~\ref{fig_sr1_toy}(d)-\ref{fig_sr1_toy}(f).
Still, the common feature of both models is that the position of the peak gets closer to $\alpha_{\rm c}$ when $q$ is increased.
All these results confirm that the similarity of the critical behavior in both models also emerges on a level of entanglement entropies.

So far, we mostly focused on the $q$ dependence of ${\mathcal S}^{(q)}$ and its impact on the determination of the critical point.
Results in Figs.~\ref{fig_sr1_toy} and~\ref{fig_sr1_sun} showed that one can pinpoint very accurately the critical point using ${\mathcal S}^{(q)}$ at $p=1$, i.e., when studying the entanglement properties of the subsystem that consists of a single, most distant spin-1/2 particle from the ergodic quantum dot.
A natural question is then to ask what is the optimal size of the subsystem that is most sensitive to the emergence of the critical behavior.
In Figs.~\ref{fig_sr2_toy} and~\ref{fig_sr2_sun} of Appendix~\ref{sec:appendix_Sr} we show results that are analogous to those in Figs.~\ref{fig_sr1_toy} and~\ref{fig_sr1_sun}, respectively.
However, they are calculated for the subsystems that consist of $p=4$ most distant particles from the dot.
Results in Figs.~\ref{fig_sr2_toy} and~\ref{fig_sr2_sun} suggest that they are still consistent with the emergence of the critical point in the vicinity of $\alpha=\alpha_{\rm c}$, nevertheless, the accuracy is not as good as for the results at $p=1$, in particular for the QSM.
From these we conclude that, at least for the QSM, the most valuable information about the transition is encoded in the particles that are most distant from the ergodic quantum dot.
In other words, when the ergodic bubble fails to thermalize the entire system, these particles are the first to exhibit nonergodic properties.

\subsection{Schmidt gaps} \label{sec:Schmidt}

\begin{figure}[!b]
\centering
\includegraphics[width=0.48
\textwidth]{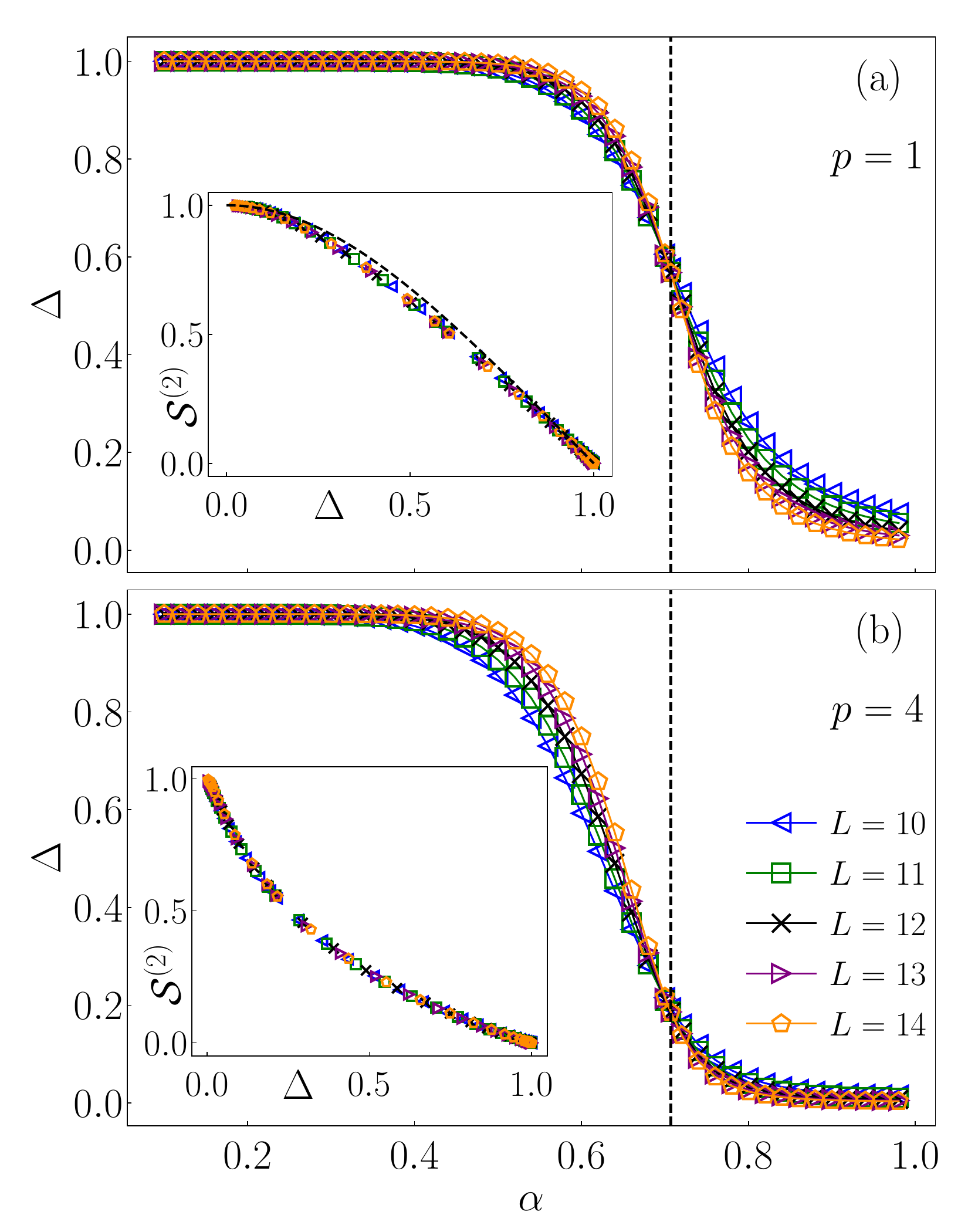}
\caption{
Schmidt gap $\Delta$ in the UM at $J=1$ and $N=1$.
(a) $p=1$, (b) $p=4$.
The main panels show $\Delta$ versus $\alpha$ at different $L$,
while the insets show the second Rényi entanglement entropy ${\cal S}^{(2)}$ from Eq.~(\ref{eq:def_renyi}) versus $\Delta$, also at different $L$.
Vertical dashed lines in the main panels denote $\alpha=\alpha_{\rm c}$ from Eq.~(\ref{eq:alpha_c}), and the dashed line in the inset of (a) is the result from Eq.~(\ref{eq:Sq_vs_Delta}).
Results for $\Delta$ are averaged over $N_\mathrm{samples}=1000,\, 400, \,300$ Hamiltonian realizations for $L\leq 12\,, L=13, \, L=14, $ respectively. 
Averaging over the eigenstates, $\langle \cdots \rangle_n,$ was performed over $N_{\mathrm{eig}} = 1000$ eigenstates near the center of the spectrum for each data point.
}
\label{fig_Sgap_toy}
\end{figure}

\begin{figure}[h]
\centering
\includegraphics[width=0.48\textwidth]{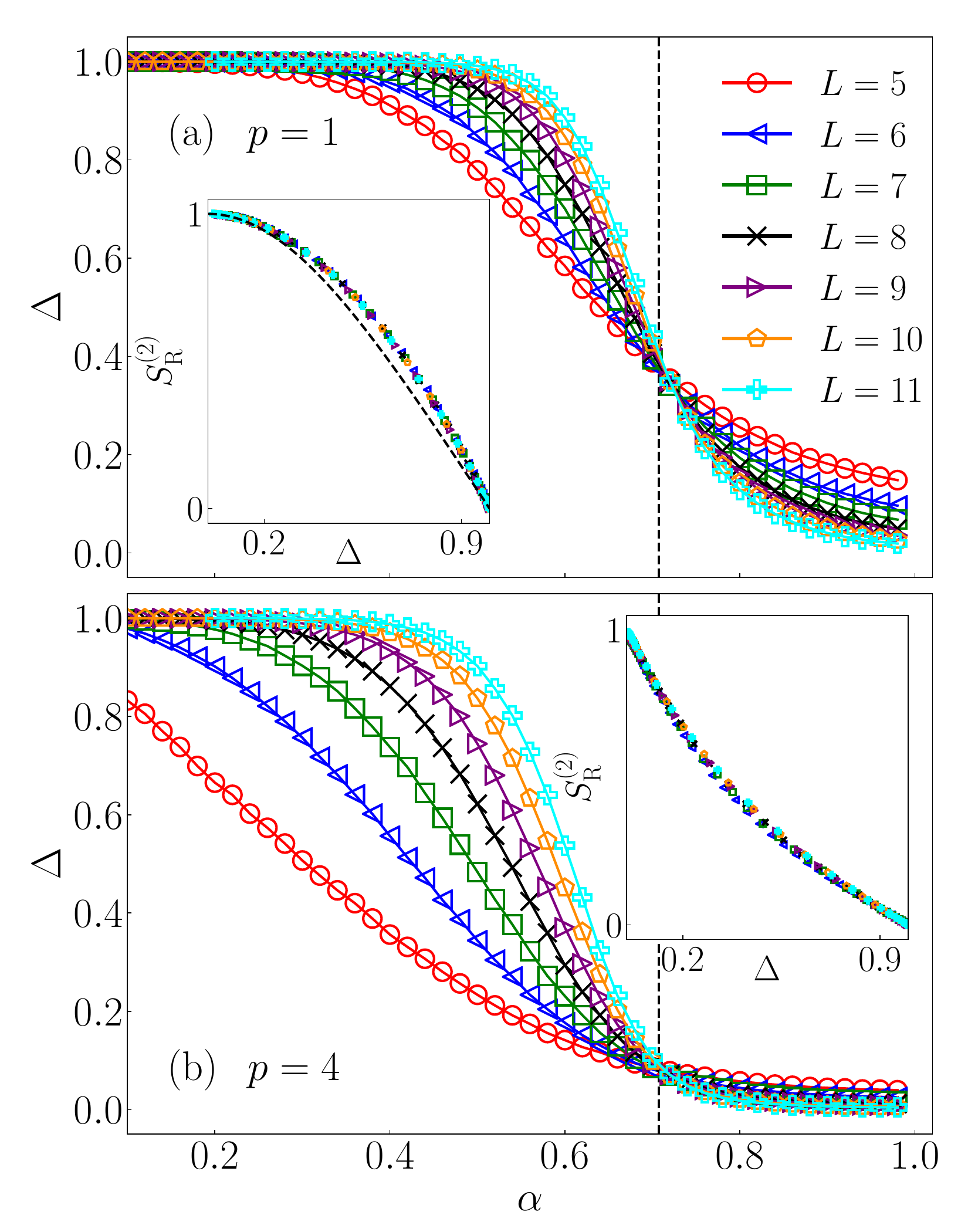}
\caption{
Schmidt gap $\Delta$ in the QSM at $g_0=\gamma=1$ and $N=5$.
(a) $p=1$, (b) $p=4$.
The main panels show $\Delta$ versus $\alpha$ at different $L$,
while the insets show the second Rényi entanglement entropy ${\cal S}^{(2)}$ from Eq.~(\ref{eq:def_renyi}) versus $\Delta$, also at different $L$.
Vertical dashed lines in the main panels denote $\alpha=\alpha_{\rm c}$ from Eq.~(\ref{eq:alpha_c}), and the dashed line in the inset of (a) is the result from Eq.~(\ref{eq:Sq_vs_Delta}).
Results for $\Delta$ are averaged over $N_\mathrm{samples}=500$ Hamiltonian realizations.
Averaging over the eigenstates, $\langle \cdots \rangle_n,$ was performed over $N_{\mathrm{eig}} = 500$ eigenstates near the center of the spectrum for each data point.
}
\label{fig_Sgap_sun}
\end{figure}

As another indicator of the critical point, which is related to the entanglement entropies, we consider the Schmidt gap $\Delta$.
The latter is defined as the difference between the two largest eigenvalues of the reduced density matrices of the subsystem $B$ from Eq.~(\ref{eq:def_rdm}), $\Delta = \lambda_1 - \lambda_2$.
In the actual numerical calculations, we then average $\Delta$ over the midspectrum eigenstates and over different Hamiltonian realizations,
\begin{equation} \label{eq:def_Sgap}
    \Delta = \langle \langle \lambda_1 - \lambda_2 \rangle_n \rangle_H \;,
\end{equation}
similar to the other quantities studied before.
Also similar to the other quantities, in $\Delta$ we omit the index $p$ indicating the number of spin-1/2 particles in the subsystem.

As already discussed in the previous section, it is convenient to study the case of a single two-level system ($p=1$), for which $\lambda_2 = 1- \lambda_1$ and hence
\begin{equation} \label{eq:Delta_p1}
\Delta = \langle\langle 2\lambda_1 - 1 \rangle_n\rangle_H \;.
\end{equation}
Results for $\Delta$ at $p=1$ are shown for the UM and the QSM in the main panels of Figs.~\ref{fig_Sgap_toy}(a) and~\ref{fig_Sgap_sun}(a), respectively.
They provide an extremely accurate measure of the critical point.
In particular, we observe that $\Delta$ is scale invariant at $\alpha=\alpha_{\rm c}$, and it tends towards $\Delta\to 0$ at $\alpha>\alpha_{\rm c}$ and $\Delta\to 1$ at $\alpha<\alpha_{\rm c}$ in the thermodynamic limit $L\to\infty$.
This establishes the Schmidt gap $\Delta$ as a candidate that may play a role of an order parameter for the ergodicity breaking phase transition.

A natural question is to ask about the optimal size of the subsystem that exhibits most sharp signatures of the transition.
To this end we extend the analysis of $\Delta$ to $p=4$ in Figs.~\ref{fig_Sgap_toy}(b) and~\ref{fig_Sgap_sun}(b), i.e., to the case where the subsystem consists of four most distant particles from the ergodic quantum dot.
Also in this case, one can still detect the critical point by inspecting the position of the scale invariant point of $\Delta$.
However, in this case the scale invariant value of $\Delta$ is much closer to zero, and hence the signatures of the scale invariance are not as sharp as in the case of $p=1$.

At $p=1$, both the Schmidt gap $\Delta$ and the entanglement entropies ${\mathcal S}^{(q)}$ are functions of a single eigenvalue $\lambda_1$, as expressed by Eqs.~(\ref{eq:Delta_p1}) and~(\ref{eq:S_p1}), respectively.
It is then expected that the ${\mathcal S}^{(q)}$ is a well-defined function of $\Delta$, independent of the system size $L$.
This property is demonstrated in the insets of Figs.~\ref{fig_Sgap_toy}(a) and~\ref{fig_Sgap_sun}(a) at $q=2$.
An interesting detail of these figures, though, is that the results do not exactly follow the expected behavior
\begin{equation} \label{eq:Sq_vs_Delta}
    {\mathcal S}^{(2)} = 1 - \log_2(1+\Delta^2) \;,
\end{equation}
that would have been derived from Eqs.~(\ref{eq:Delta_p1}) and~(\ref{eq:S_p1}) if there was no averaging over eigenstates and Hamiltonian realizations.
It then appears to be not entirely trivial that at $p=1$, ${\mathcal S}^{(2)}$ is a well-defined function of $\Delta$ despite this function being different from Eq.~(\ref{eq:Sq_vs_Delta}).
As a side remark, in Fig.~\ref{fig_sGap_toy_2}(a) of Appendix~\ref{sec:appendix_Sgap} we show that when no averaging over Hamiltonian eigenstates and Hamiltonian realizations are performed, the second Rényi entanglement entropy is indeed a well-defined function of the Schmidt gap as predicted by Eq.~(\ref{eq:Sq_vs_Delta}). 
The deviations between Eq.~(\ref{eq:Sq_vs_Delta}) and the numerical results in the insets of Figs.~\ref{fig_Sgap_toy}(a) and~\ref{fig_Sgap_sun}(a) are hence a consequence of the averaging.

At $p>1$, we are not aware of any formal argument to predict a unique relationship between ${\mathcal S}^{(q)}$ and $\Delta$.
Quite surprisingly, however, we still observe a nearly perfect collapse of the results for ${\mathcal S}^{(2)}$ when plotted at $p=4$ as a function of $\Delta$, see the insets of Figs.~\ref{fig_Sgap_toy}(b) and~\ref{fig_Sgap_sun}(b).
A detailed inspection in Fig.~\ref{fig_sGap_toy_2} of Appendix~\ref{sec:appendix_Sgap} reveals that the collapse is not present when ${\mathcal S}^{(2)}$ is plotted versus $\Delta$ for a single Hamiltonian eigenstate, and that the signatures of a well-defined functional dependence of ${\mathcal S}^{(2)}$ on $\Delta$ can be readily observed after the averaging over eigenstates within a single Hamiltonian realization.
While we are here not able to fully rationalize this behavior, we note that the remarkable scaling collapses share close similarities in both models.

\section{Multifractality and Fock space localization} 
\label{sec:multifractality}

Having established in Sec.~\ref{sec:indicators} the sharp agreement between the numerically determined location of the critical point and the analytical prediction, we here focus on the properties on the nonergodic side and at the critical point.
The central quantity of study will be the fractal dimension introduced below.

\subsection{Fractal dimension} \label{sec:fractal}

In this work we extract the fractal dimension from the participation entropies, which were introduced in Sec.~\ref{sec:ipr}.
In particular, in Sec.~\ref{sec:ipr} we focused on the participation entropies $S_{\rm typ}^{(q)}$ that we calculated from the typical IPR, see Eq.~(\ref{eq:def_participation_entro}), and hence we refer to the corresponding fractal dimensions as $d_{\rm typ}^{(q)}$.
We calculate $d_{\rm typ}^{(q)}$ from the ansatz~\cite{Mace2019}
\begin{equation} \label{eq:def_Sq_dq}
S_{\rm typ}^{(q)} = d_{\rm typ}^{(q)} \ln{\mc D} + b_{\rm typ}^{(q)}\;,
\end{equation}
where ${\cal D} = 2^{N+L}$ is the Fock space dimension and $b_{\rm typ}^{(q)}$ is an $L$-independent constant.
The ansatz in Eq.~(\ref{eq:def_Sq_dq}) is phenomenological, and it is expected to describe well the results in the asymptotic regime, i.e., at sufficiently large $L$ (the details of the numerical implementation will be discussed below).
The fractal dimension $d_{\rm typ}^{(q)}$ is an $L$-independent constant that may depended on $q$.
If that is the case, we refer to the wavefunction properties as being multifractal.
We note that the fractal dimension is a basis-dependent quantity since it depends on the wavefunction coefficients $c_i = \braket{i | n}$ calculated in some basis $\{|i\rangle\}$.
Here, $\{|i\rangle\}$ corresponds to the computational basis.

A similar but nonequivalent way to define the fractal dimension is via the average IPR, which we denote as $\langle \langle P_q^{-1} \rangle_n\rangle_{H}$.
In this case, one extracts the decay coefficient $\tau_{}^{(q)}$ of the IPR from the ansatz
\begin{equation}
\langle \langle P_q^{-1} \rangle_n\rangle_{H} \propto {\mc D}^{-\tau_{}^{(q)}} \;,
\end{equation}
from which one can define the fractal dimension $d^{(q)}$ via the relation $d_{}^{(q)}=\tau_{}^{(q)}/(q-1)$.
Nevertheless, studies of single-particle Anderson localization transition noted that the distribution of the IPRs at the critical point may be broad~\cite{Evers2000}, and hence a more natural choice of consideration is $d_{\rm typ}^{(q)}$ instead of $d^{(q)}$.
In all numerical results reported here we only focus on $d_{\rm typ}^{(q)}$.
We observe (not shown here) no significant differences between $d^{(q)}$ and $d_{\rm typ}^{(q)}$, however, a quantitative comparison between $d^{(q)}$ and $d_{\rm typ}^{(q)}$ is left for future work. 
For the sake of completeness, we also introduce the decay coefficient of the typical IPR, denoted as $\tau_{\rm typ}^{(q)}$, such that
\begin{equation}\label{eq:def_fractal_dimension}
    \tau^{(q)}_{\rm typ} = d^{(q)}_{\rm typ}(q - 1)\;.
\end{equation}
Both quantities $d^{(q)}_{\rm typ}$ and $\tau^{(q)}_{\rm typ}$ will be discussed below.

The behavior of the fractal dimension is well understood in the ergodic phase and in the nonergodic phase that exhibits Fock space localization.
These considerations apply to both $d^{(q)}$ and $d_{\rm typ}^{(q)}$~\cite{Rodriguez2011}, and for simplicity we here only discuss the former.
In the ergodic phase that exhibits many-body quantum chaos~\cite{dalessio_kafri_16}, the wavefunction coefficients $c_i = \braket{i | n}$ can be considered as normally distributed random variables with zero mean and variance $1/{\cal D}$, and hence $\langle \langle P_q^{-1} \rangle_n\rangle_{H}$ scales with the Hilbert space dimension approximately as 
\begin{equation}
\langle \langle P_q^{-1} \rangle_n\rangle_{H} \propto \sum\limits_{i=1}^{\mc D}\frac{1}{{\mc D}^q} = {\mc D}^{1-q} \;,
\end{equation}
giving rise to the decay coefficient $\tau_{}^{(q)} = q-1$ and a $q$-independent fractal dimension $d_q = 1$.
Specifically, in the commonly studied case at $q=2$, the IPR scales as $\langle \langle P_2^{-1} \rangle_n\rangle_{H} \propto {\mc D}^{-1}$, which is in agreement with the exact result predicted by the GOE of the RMT~\cite{mehta_91}, $\langle \langle P_2^{-1} \rangle_n\rangle_{H} = 3/{\mc D}.$ 
In the opposite case of Fock space localization, one may think of only ${\mc O}(1)$ coefficients $c_i$ begin nonzero, and thus $\langle \langle P_q^{-1} \rangle_n\rangle_{H}$ does not scale with the system size.
This implies $\tau_{}^{(q)} = d_{}^{(q)} = 0$ for all $q$.
The only exception is $q=1$, at which $P_q^{-1} = 1$ due to normalization, and hence $d_{}^{(1)}$ cannot distinguish between the ergodic and nonergodic phase.
However, studying the typical fractal dimension $d_{\rm typ}^{(q)}$ that is the focus of our study, no such limitation occurs at $q=1$.

\begin{figure}[!b]
\centering
\includegraphics[width=0.48\textwidth]{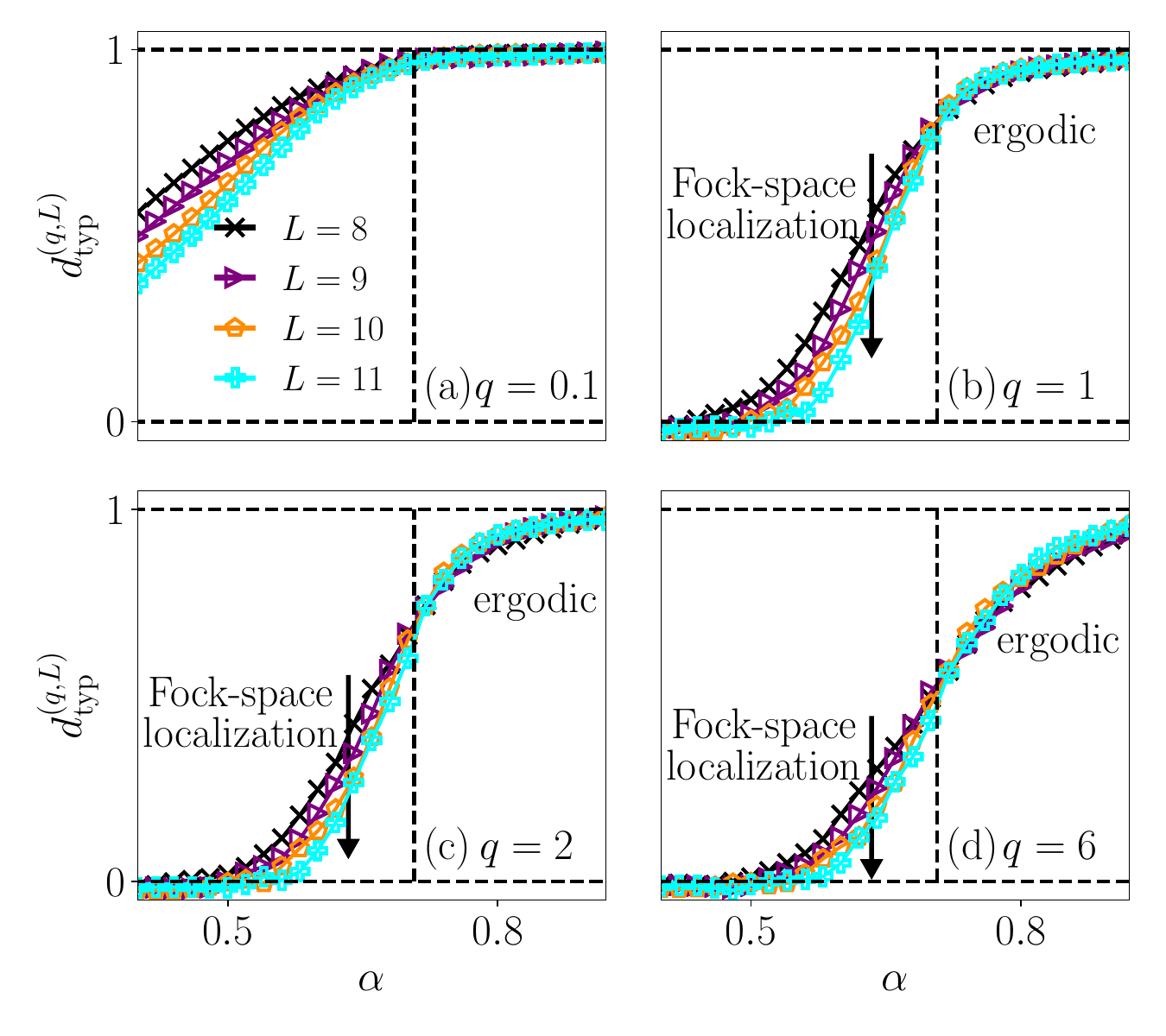}
\caption{The $L$-dependent fractal dimension $d_{\rm typ}^{(q,L)}$ from Eq.~(\ref{eq:def_dqL}) in the QSM at $g_0=\gamma=1$ and $N=5$, at (a) $q=0.1$, (b) $q=1$, (c) $q=2$ and (d) $q=6$.
The arrows denote the flows towards Fock space localization in the asymptotic regime.
Vertical dashed lines denote $\alpha=\alpha_{\rm c}$ from Eq.~(\ref{eq:alpha_c}).
}
\label{fig_tauq1_flow_sun}
\end{figure}

\begin{figure}[!t]
\centering
\includegraphics[width=0.98\columnwidth]{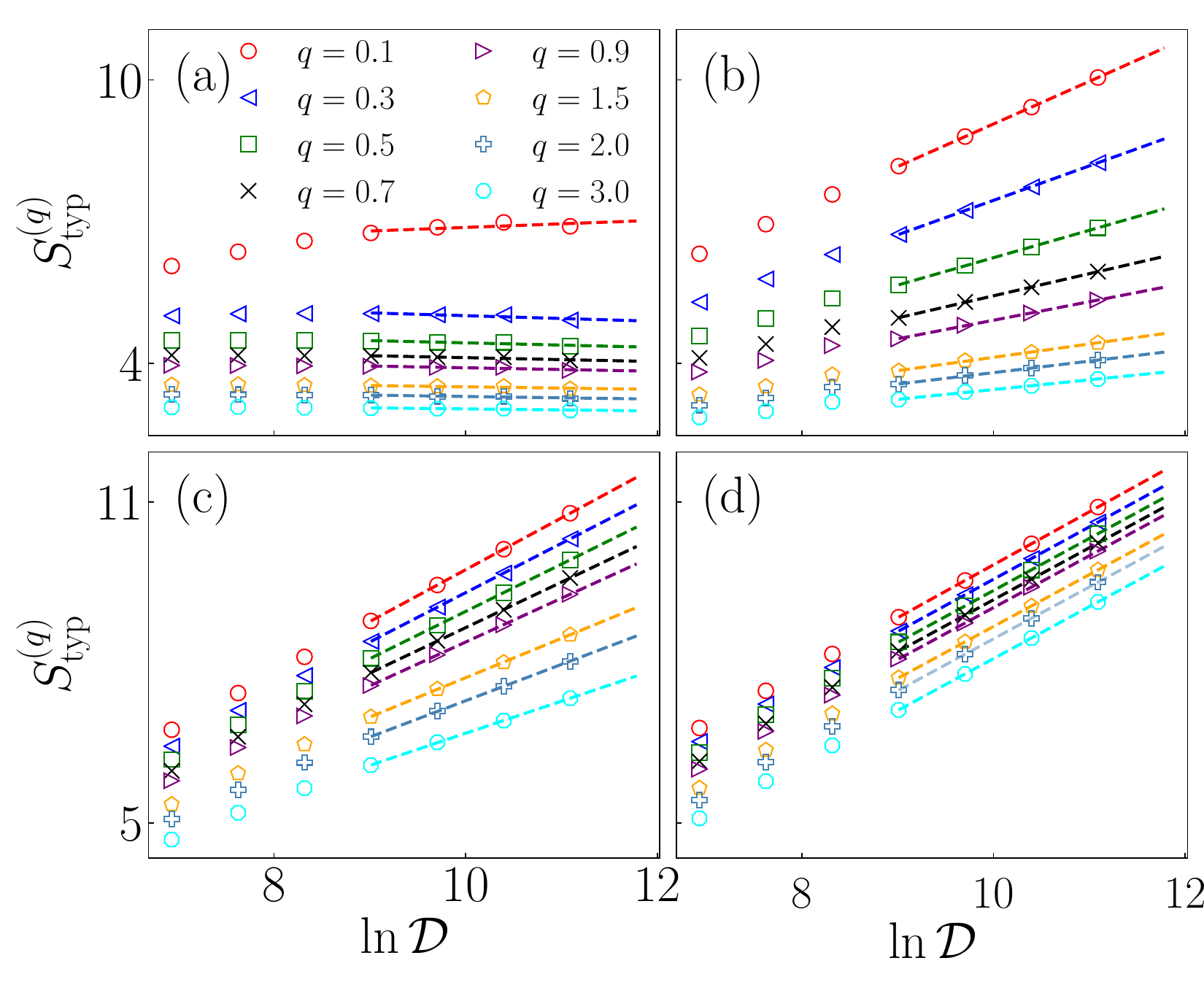}
\caption{
Participation entropy $S_{\rm typ}^{(q)}$ versus the logarithm of the Fock space dimension $\ln {\cal D}$ in the QSM at (a) $g_0=\gamma=1$, $\alpha=0.2$, (b) $g_0=0.3, \gamma=0.3$, $\alpha=\overline{\alpha_{}}=0.81$, (c) $g_0=\gamma=1$, $\alpha=\alpha_{\rm c}$ (d) $g_0=\gamma=1$, $\alpha=0.9$.
Lines are fits according to the ansatz in Eq.~(\ref{eq:def_Sq_dq}) for Fock space dimensions of the total system in the range from ${\cal D}=2^{13}$ to ${\cal D}=2^{16}$.
They allow for the extraction of the fractal dimension $d^{(q)}_\mathrm{typ}$ and the coefficient $\tau^{(q)}_\mathrm{typ}$, plotted in Fig.~\ref{fig_tauq1_sun_short}.
}
\label{fig_tauq1_sun}
\end{figure}

In finite systems such as those studied here, one often obtains $d_{\rm typ}^{(q)}$ that is away from the limiting cases discussed above, i.e., $0<d_{\rm typ}^{(q)}<1$.
This result may either be consistent with (multi)fractality of the system, or suggesting that the system is not yet in the asymptotic regime.
In our numerical analyses, we first verify whether the system can be considered as being in the asymptotic regime, and hence if the ansatz from Eq.~(\ref{eq:def_Sq_dq}) can be applied.
Figures~\ref{fig_tauq1_flow_sun},~\ref{fig_tauq1_sun} and~\ref{fig_tauq1_toy} show examples in which the system is close to the asymptotic regime: these occur deep in the ergodic regime at $\alpha_c \ll \alpha \lesssim 1$ and at the critical point $\alpha=\alpha_c$. 
In contrast,  outside the asymptotic regime, $d^{(q)}_{\rm typ}$ should be replaced by a number that depends on $L$, i.e., $d^{(q)}_{\rm typ} \rightarrow d^{(q,L)}_{\rm typ}$, which we calculate in Fig.~\ref{fig_tauq1_flow_sun}.
We numerically extract $d^{(q,L)}_{\rm typ}$ by computing the values of slope of $S_{\rm typ}^{(q)}$ between consecutive system sizes $L-1$ and $L$, 
\begin{equation} \label{eq:def_dqL}
d^{(q,L)}_{\rm typ}= \frac{S_{\rm typ}^{(q)}(L)-S_{\rm typ}^{(q)}(L-1)}{\ln[{\mc D}(L)]-\ln[{\mc D}(L-1)]}\;.
\end{equation}
Such procedure was recently used in studies of multi-fractal properties of wavefunctions on different types of Anderson graphs~\cite{sierant_lewenstein_23, GarciaMata2020, GarciaMata2022}.
From the flow of $d^{(q,L)}_{\rm typ}$ with $L$, one may conjecture about the fate of $d^{(q)}_{\rm typ}$ in the thermodynamic limit $L\to \infty$.

\subsection{Fock space localization on the nonergodic side}

We first focus on the properties on the nonergodic side of the critical point, at $\alpha < \alpha_c$.
Previous studies in the UM have established, almost up to a rigorous level, that in the nonergodic phase the system exhibits Fock space localization~\cite{fyodorov2009anderson,vansoosten2018phase}.

\begin{figure*}[!t]
\centering
\includegraphics[width=0.95\textwidth]{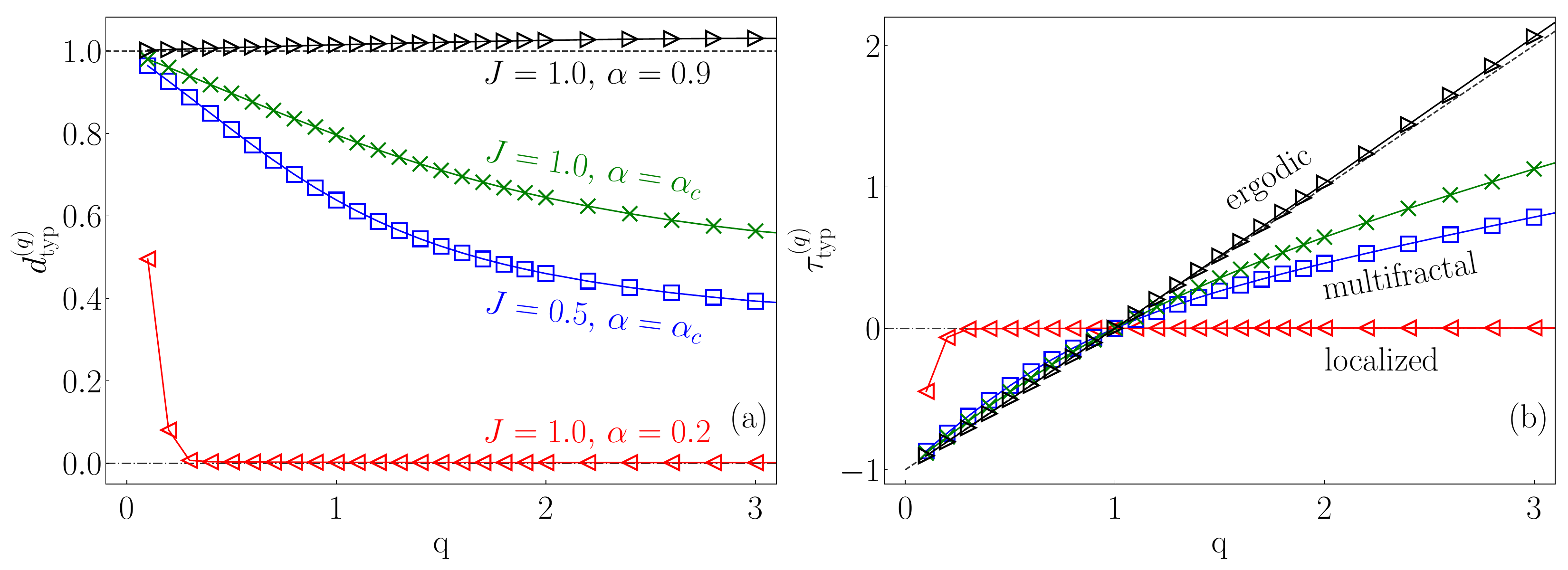}
\caption{
(a) Fractal dimension $d^{(q)}_{\rm typ}$ versus $q$, and (b) the decay coefficient $\tau^{(q)}_{\rm typ}$ of the typical IPR versus $q$, in the UM at $N=1$.
The values are extracted from the $S_{\rm typ}^{(q)}$ versus $\ln {\cal D}$ curves, such as those shown in Fig.~\ref{fig_tauq1_toy}, using the ansatz in Eq.~(\ref{eq:def_Sq_dq}).
The dashed lines in both panels correspond to the ergodic limit $d^{(q)}_{\rm typ}=1$, $\tau^{(q)}_{\rm typ} = q-1$, and the dashed-dotted lines correspond to the nonergodic limit $d^{(q)}_{\rm typ}=\tau^{(q)}_{\rm typ}=0$.
}
\label{fig_tauq1_toy_short}
\end{figure*}

\begin{figure*}[!t]
\centering
\includegraphics[width=0.95\textwidth]{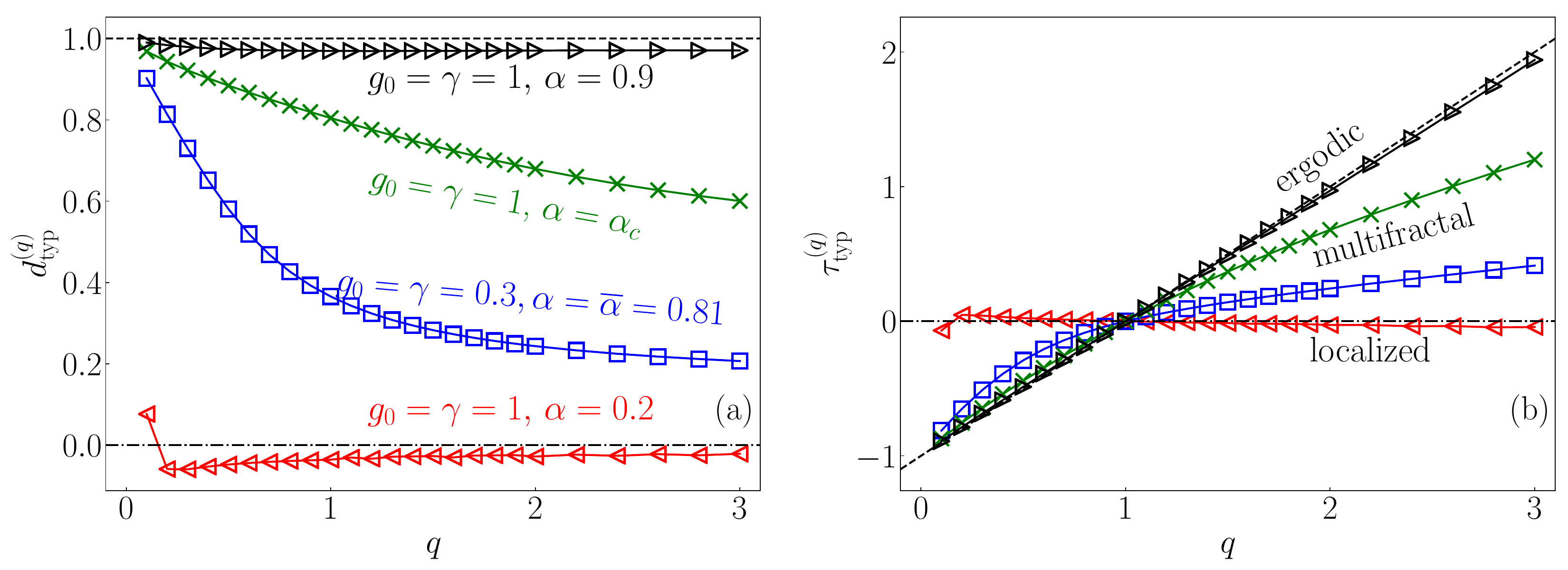}
\caption{
(a) Fractal dimension $d^{(q)}_{\rm typ}$ versus $q$, and (b) the decay coefficient $\tau^{(q)}_{\rm typ}$ of the typical IPR versus $q$, in the QSM at $N=5$.
The values are extracted from the $S_{\rm typ}^{(q)}$ versus $\ln {\cal D}$ curves, such as those shown in Fig.~\ref{fig_tauq1_sun}, using the ansatz in Eq.~(\ref{eq:def_Sq_dq}).
The dashed lines in both panels correspond to the ergodic limit $d^{(q)}_{\rm typ}=1$, $\tau^{(q)}_{\rm typ} = q-1$, and the dashed-dotted lines correspond to the nonergodic limit $d^{(q)}_{\rm typ}=\tau^{(q)}_{\rm typ}=0$.
}
\label{fig_tauq1_sun_short}
\end{figure*}

Here we are particularly interested on the nonergodic side of the QSM.
The latter can be divided in two regimes: 
the regime close to the critical point and the regime deep in the nonergodic phase.
Figure~\ref{fig_tauq1_flow_sun} shows the $L$-dependent fractal dimension $d_{\rm typ}^{(q,L)}$ from Eq.~(\ref{eq:def_dqL}), which suggest that at sufficiently large $q \gtrsim 1$, the first regime approximately belongs for the interval $0.5 \lesssim \alpha < \alpha_c$.
In contrast, at $\alpha \lesssim 0.5$ and $q\gtrsim 1$ the system exhibits clear signatures of Fock space localization since $d_{\rm typ}^{(q)} \approx 0$.
These results are corroborated in Figs.~\ref{fig_tauq1_toy_short} and~\ref{fig_tauq1_sun_short} by the results at $\alpha=0.2$ in the UM and the QSM, respectively, which consistently show $d_{\rm typ}^{(q)} \approx \tau_{\rm typ}^{(q)} \approx 0$.
In fact, we even observe slightly negative values of $d_{\rm typ}^{(q)}$, see the results at $\alpha=0.2$ in Figs.~\ref{fig_tauq1_sun}(a) and~\ref{fig_tauq1_sun_short}.
In this case, we expect $d_{\rm typ}^{(q)} \to 0$ in the thermodynamic limit.

The limit $q\to 0$, which is not the focus of this study, may represent an exception to these considerations.
As discussed in Sec.~\ref{sec:models_um}, it is understood from the studies of the PLRBM~\cite{mirlin_fyodorov_96,EversMirlin2008} that the observation of $L$-independent IPR (and consequently the participation entropies) may only emerge deep in the localized regime, i.e.\ for sufficiently large $a$, with the threshold value for $a$ depending on $q$.
Similar arguments likely apply to the UM, and it is beyond the scope of this paper to establish to what degree these arguments also apply to the QSM.

The most interesting aspect of Fig.~\ref{fig_tauq1_flow_sun} is the flow of $d^{(q,L)}_{\rm typ}$ upon increasing $L$.
In the nonergodic regime at $0.5 \lesssim \alpha < \alpha_c$, the results are consistent with a flow towards Fock space localization in the thermodynamic limit, i.e., $d^{(q,L)}_{\rm typ} \to 0$ when $L \to \infty$.
These results are also consistent with the recent analysis of $d^{(q)}$ at $\alpha=0.6$ and $q=2$ in the QSM~\cite{hopjan2023}, which was interpreted as $d^{(q)}\to 0$.

On the other hand, the results at $\alpha>\alpha_c$ in Fig.~\ref{fig_tauq1_flow_sun} exhibit values that are close to, but not necessary equal to $d^{(q,L)}_{\rm typ} = 1$.
In the vicinity of the critical point, $d^{(q,L)}_{\rm typ}$ is nonzero and lower than 1, which we interpret as a signature of multifractality and will be discussed in more details in the next section.

Summarizing these results, we expect that Fock space localization is a property of the entire nonergodic phase in both the UM and the QSM.
This is different from what is expected to occur in the putative MBL phase, for which emergence of multifractality was conjectured, based on simple theoretical arguments, in the entire phase~\cite{Mace2019,Luitz2020,Roy2021,detomasi_khaymovich_21,Orito2021,Tarzia2020}.

\subsection{Multifractality on the critical manifold}

We now focus our attention on the eigenstate wavefunction properties at the critical point.
In the UM it is understood that the wavefunction is multifractal \cite{rushkin2011universal,MendezBermudez2012,Bogomolny2018}.
Specifically, it was shown that $d^{(q)}$ exhibits $q$ dependence and in the limit $J\to 0$, explicit expressions were derived for $d^{(q)}$. In lowest order in $J$, the $q$ dependence of $d^{(q)}$ is the same for the UM and the PLRBM~\cite{mirlin_fyodorov_96, rushkin2011universal}.

Beyond the UM, multifractality was observed in several random matrix ensembles~\cite{Fyodorov1991,Evers2000,Pino2019,Kravtsov2015}.
Its concept was also extended to physical models without interactions such as Anderson models on hypercubic lattices~\cite{EversMirlin2008, Rodriguez2009, Rodriguez2010, Rodriguez2011, Devakul2017, Tarquini2017} and on graphs~\cite{DeLuca2014, sierant_lewenstein_23, GarciaMata2022,GarciaMata2020,Pino2017,GarciaMata2017}, and the connections were made to quantum dynamics~\cite{Ketzmerick1997, Ohtsuki1997, Kravtsov2011, DeTomasi2019, Bera2018,Kravtsov2010,Ng2006, Torres-Herrera2015, hopjan2023}.
A currently very active direction of research is to explore to which degree the concept of multifractality can be applied to many-body quantum systems at the boundary of quantum chaos, such as those studied in the context of MBL~\cite{Mace2019,Roy2021,detomasi_khaymovich_21,Orito2021,Luitz2020,Tarzia2020}.

Here we first complement previous results on multifractality in the UM by showing that also $d_{\rm typ}^{(q)}$ exhibits multifractal properties.
Specifically, in Figs.~\ref{fig_tauq1_toy}(b) and~\ref{fig_tauq1_toy}(c) in Appendix~\ref{sec:appendix_Fractal} we show that $S_{\rm typ}^{(q)}$ can be well fitted by the ansatz from Eq.~(\ref{eq:def_Sq_dq}), suggesting that the system is indeed very close to the asymptotic regime.
We then show the extracted values of $d_{\rm typ}^{(q)}$ and $\tau_{\rm typ}^{(q)}$ as a function of $q$ in Fig.~\ref{fig_tauq1_toy_short}.
The fractal dimension $d_{\rm typ}^{(q)}$ decreases with $q$, and this dependence on $q$ confirms the multifractal character of the wavefunction.

Figure~\ref{fig_tauq1_toy_short} actually shows $d_{\rm typ}^{(q)}$ and $\tau_{\rm typ}^{(q)}$ versus $q$ for two different parameter values $J=1$ and $J=0.5$ at the critical point $\alpha=\alpha_c$.
Different values of $J$ give rise to different fractal dimensions of wavefunctions, suggesting the emergence of a manifold of critical points.
In Sec.~\ref{sec:spectrum_critical} we further study the impact of $J$ on the critical properties, specifically, on the spectral statistics.
It is possible that there is a one-parameter family of critical points tuned by the parameter $J$, however, we cannot rule out the existence of a higher-dimensional critical manifold.

\begin{figure*}[!t]
\centering
\includegraphics[width=0.99\textwidth]{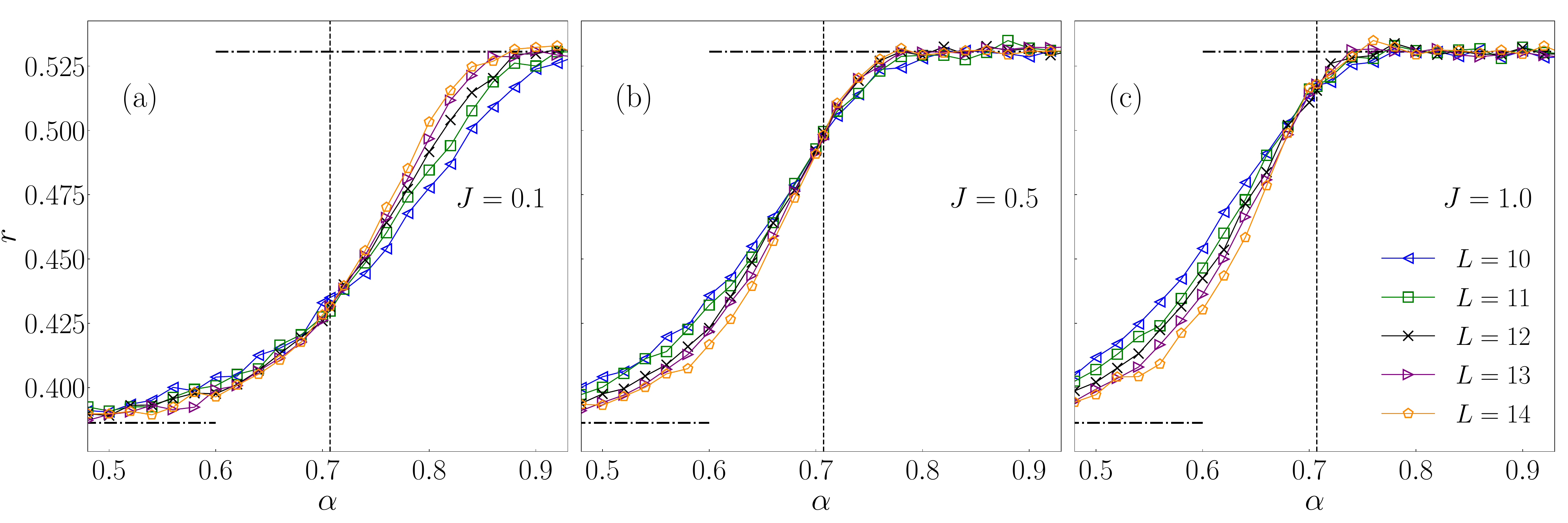}
\caption{
The average gap ratio $r$ defined in Eq.~(\ref{eq:def_r}) in the UM at $N=1$, as a function of $\alpha$ and for different system sizes $L$.
(a) $J=0.1$, (b) $J=0.5$, (c) $J=1$.
The vertical dashed lines denote the prediction for the critical point $\alpha=\alpha_c$ from Eq.~(\ref{eq:alpha_c}).
The GOE and Poisson limits for $r$ are denoted by the upper and lower horizontal dashed-dotted lines, respectively. 
}
\label{fig_r2_toy}
\end{figure*}

\begin{figure}[!t]
\centering
\includegraphics[width=0.460\textwidth]{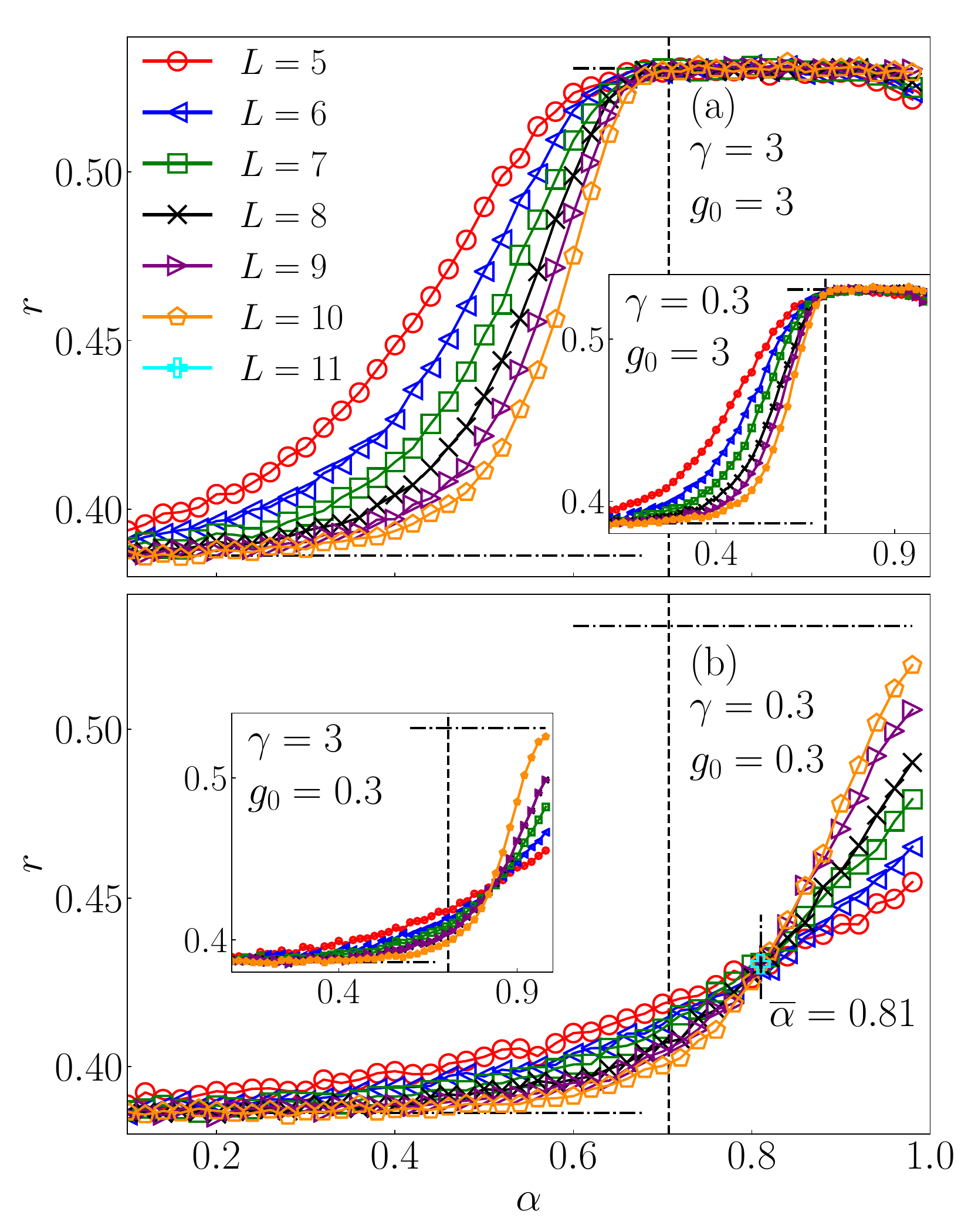}
\caption{
The average gap ratio $r$ defined in Eq.~(\ref{eq:def_r}) in the QSM at $N=5$, as a function of $\alpha$ and for different system sizes $L$.
(a) $g_0=3$, with $\gamma=3$ in the main panel and $\gamma=0.3$ in the inset.
(b) $g_0=0.3$, with $\gamma=0.3$ in the main panel and $\gamma=3$ in the inset.
The long vertical dashed lines denote the prediction for the critical point $\alpha=\alpha_c$ from Eq.~(\ref{eq:alpha_c}), while the short vertical line in the main panel of (b) denotes the crossing point at $\overline{\alpha}=0.81$.
The GOE and Poisson limits for $r$ are denoted by the upper and lower horizontal dashed-dotted lines, respectively.
}
\label{fig_r2_sun}
\end{figure}

The central question is whether the QSM also exhibits multifractal properties at the critical point.
If the answer is affirmative, the next question is then to explore whether one can tune the multifractal properties within the critical manifold as in the UM.

The results in Fig.~\ref{fig_tauq1_flow_sun} have already suggested that the $L$-dependent fractal dimension $d_{\rm typ}^{(q,L)}$ exhibits no, or only mild, dependence on $L$ at the critical point of the QSM.
This observation is confirmed by the results in Figs.~\ref{fig_tauq1_sun}(b) and~\ref{fig_tauq1_sun}(c), which exhibit a linear dependence of $S_{\rm typ}^{(q)}$ on $\ln{\cal D}$ according to Eq.~(\ref{eq:def_Sq_dq}), and hence suggest that the system is already very close to the asymptotic regime in which the $L$-dependence on $d_{\rm typ}^{(q)}$ should not be large.

The results for the fractal dimension $d^{(q)}_{\rm typ}$, and for the decay coefficient $\tau^{(q)}_{\rm typ}$ of the typical IPR, are shown in Fig.~\ref{fig_tauq1_sun_short}.
At the critical point, they exhibit two important features: $d^{(q)}_{\rm typ}$ is nonzero and lower than 1 for all nonzero values of $q$ under consideration, and it exhibits clear $q$ dependence (it monotonically decreases with $q$).
While the first property may be argued as being consistent with the results of Ref.~\cite{hopjan2023}, the second property of the QSM has to our knowledge not yet been explored.
We interpret these results as evidence of the multifractal character of the wavefunctions.

The results for the QSM in Fig.~\ref{fig_tauq1_sun_short} appear to be very similar to those for the UM in Fig.~\ref{fig_tauq1_toy_short}.
Interestingly, one also observes similarity when studying the role of the coupling parameters $J$ and $g_0$ in the UM and the QSM, respectively.
In both cases, tuning the parameters $J$ and $g_0$ from large to small values gives rise to a decrease of $d^{(q)}_{\rm typ}$, which we interpret as tuning the multifractality from weak to strong.

Note that the results in Sec.~\ref{sec:indicators} have established that the location of the critical point in the QSM is sharply predicted by $\alpha=\alpha_c$ from Eq.~(\ref{eq:alpha_c}) when all the model parameters are of the same order, which is certainly the case at $g_0=\gamma=1$.
If, however, the coupling parameter $g_0$ is varied, which is going to be studied in more detail in Sec.~\ref{sec:spectrum_critical}, the transition point observed in finite systems may not accurately coincide with $\alpha_c$.
For example, at $g_0=0.3$ that is also studied in Fig.~\ref{fig_tauq1_sun_short}, the transition point from the $r$ statistics appears to be located around $\overline{\alpha}=0.81$, see Fig.~\ref{fig_r2_sun}(b).
This is the reason why multifractal properties of the QSM in Fig.~\ref{fig_tauq1_sun_short} are, at $g_0=0.3$, studied at $\alpha=\overline{\alpha}$ and not at $\alpha=\alpha_c$.

\section{Spectral statistics on the critical manifold} \label{sec:spectrum_critical}

Having established multifractal properties at the critical point of both models, we now turn our attention to the spectral properties that have already been studied in Sec.~\ref{sec:gap}.
We are particularly interested in how the spectral properties at the critical point change when the system is tuned from weak multifractality, i.e., from large fractal dimension $d_{\rm typ}^{(q)}$ (at $q \approx 1$) towards strong multifractality, i.e., towards small fractal dimension $d_{\rm typ}^{(q)}$.

In Figs.~\ref{fig_r2_toy} and~\ref{fig_r2_sun} we study the behavior of the average gap ratio $r$, defined in Eq.~(\ref{eq:def_r}), around the critical point at different values of model parameters.
We focus on the value of $r$ at the critical point, which is detected by the scale invariant crossing point of $r$ versus $\alpha$.
While in the UM the crossing point occurs at the predicted value $\alpha=\alpha_c$ for essentially all model parameters of investigation, we already highlighted in the previous sections that this may not necessary be the case in the QSM.
An example of the latter is given in Fig.~\ref{fig_r2_sun}(b), where the crossing point at $g_0=0.3$ emerges close to $\alpha=\overline{\alpha}=0.81$.
Clarifying the fate of this crossing point in the thermodynamic limit appears to be a challenging task that is beyond the scope of the paper.
We note that at this point we are not aware of any rigorous argument that would prevent the crossing point $\overline{\alpha}$ from drifting towards the predicted critical point $\alpha_c$ from Eq.~(\ref{eq:alpha_c}).

Figure~\ref{fig_r2_toy} studies the impact of the coupling parameter $J$ in the UM on the nature of the critical point.
From Fig.~\ref{fig_tauq1_toy_short} we have already learned that decreasing $J$ gives rise to smaller values of the fractal dimension $d_{\rm typ}^{(q)}$ and hence to stronger multifractality.
Figure~\ref{fig_r2_toy} suggests that stronger multifractality is associated with a smaller value of $r$ at the critical point, i.e., with spectral statistics that is closer to the Poisson distribution.
We hence expect that $J$ represents the tuning parameter of the spectral statistics within the critical manifold, spanning from the RMT-like statistics at weak multifractality to Poisson-like statistics at strong multifractality.
These properties share some analogies with the PLRBM~\cite{mirlin_fyodorov_96}, in which the characteristic length $b$ tunes the properties of the critical point, ranging from Poisson-like at strong multifractality ($b \ll 1$) to RMT-like at weak multifractality ($b \gg 1$)~\cite{mirlin_evers_00,hopjan2023a}.

Similar behaviour is observed in the QSM as a function of the coupling parameter $g_0$, see Fig.~\ref{fig_r2_sun}.
In particular, large $g_0 > 1$ drive the critical point towards RMT-like statistics, see Fig.~\ref{fig_r2_sun}(a), while small $g_0 < 1$ bring it closer to the Poission-like statistics, see Fig.~\ref{fig_r2_sun}(b).
The latter is consistent with strong multifractality observed in Fig.~\ref{fig_tauq1_sun_short} at $g_0 = 0.3$.
On the other hand, the change of $\gamma$, which governs the spectral width of the ergodic quantum dot in the QSM, does not appear to have any significant impact on the spectral statistics within the critical manifold, see the insets of Fig.~\ref{fig_r2_sun}. 

\section{Conclusion} \label{sec:conclusions}

This work establishes a direct connection of the critical behavior between the two different models, the UM of the RMT and the QSM.
A convenient aspect of both models is that there exist quantitative analytical arguments for the value of the critical point of the ergodicity breaking phase transition.
Carrying out exact numerical calculations for various ergodicity indicators, we showed that these analytical arguments sharply predict the location of the critical point.

While our main goal was to establish the similarity of the two models, we also introduced certain ergodicity measures that have previously not received much attention.
Among those, we stress the entanglement entropy of the most distant (i.e., most weakly coupled) spin-1/2 particle, which exhibits scale invariant behavior at the critical point, and the derivatives of the participation and entanglement entropies, which exhibit a sharp peak at the critical point.
Another feature of the numerical calculations is that the location of the critical point in the QSM is closest to the predicted analytical value when the critical properties comply with the RMT predictions, and it exhibits small deviations otherwise.
While this feature was already noticed before~\cite{suntajs_vidmar_22, pawlik_sierant_23,hopjan2023}, future work should explore in more details  the fate of these small deviations in the thermodynamic limit.

On the nonergodic side of the transition, we argued that both models exhibit Fock space localization in the eigenbasis of the $\hat S^z$ operator. 
While the latter was already established in the UM~\cite{fyodorov2009anderson,vansoosten2018phase}, our results suggest that also in the QSM, Fock space localization  may emerge in the entire nonergodic phase in the thermodynamic limit.
This result is different from what was proposed for the putative MBL phase in random-field spin-1/2 chains~\cite{Mace2019,Roy2021,detomasi_khaymovich_21,Orito2021,Luitz2020,Tarzia2020}, in which the entire nonergodic phase was conjectured to be multifractal.
Therefore, while one can draw certain parallels between the ergodicity breaking in the QSM and MBL, there are also clear differences between these two phenomena.

At criticality, the eigenvectors exhibit multifractal behaviour in the eigenbasis of the $\hat S^z$ operator, characterized by a family of nonzero fractal dimensions that are lower than unity.
The fractal dimensions may vary with the size of the initial ergodic seed and the overall coupling of the ergodic seed to the remainder of the system, thereby suggesting the emergence of a manifold of critical points.
This may carry some similarities with certain RMT-based models for Anderson localization transition, such as the power-law random banded matrix models~\cite{mirlin_fyodorov_96} that exhibit a one-parameter family of critical points.
However, whether there is a one-parameter family, or eventually a higher manifold of critical points in the QSM, is an open question that should be addressed in the future.

As a consequence of the emergence of the manifold of critical points, the spectral statistics on the critical manifold may be continuously varied from the RMT-like statistics to the Poisson-like statistics.
In the former, the fractal dimensions are close to unity, while in the latter they are close to zero.
In the context of single-particle transitions, such as those in the power-law random banded matrices, this relationship is in certain limits understood even analytically~\cite{mirlin_evers_00}, while a systematic study in many-body quantum systems appears to be a natural next step of investigation.

To summarize, our results reinforce the QSM and the UM as fertile playgrounds to study many-body ergodicity breaking phase transitions, and call for further characterization of their properties both from the numerical as well as from the analytical side.

\acknowledgements
We acknowledge discussions with I. M. Khaymovich.
This work is supported by the Slovenian Research and Innovation Agency (ARIS), Research core funding numbers P1-0044, N1-0273, J1-50005, and N1-0369 (J.\v S., M.H. and L.V.). 
W.D.R. was supported in part by the FWO through grant G098919N, by an internal KULeuven grant  iBOF DOA/20/011 and by the Excellence of Science (EOS) programme (FWO and  F.R.S.-FNRS) through grant EOS G0H1122N EOS 40007526 CHEQS.
We gratefully acknowledge the High Performance Computing Research Infrastructure Eastern Region (HCP RIVR) consortium~\footnote{\href{https://www.hpc-rivr.si/}{www.hpc-rivr.si}} and European High Performance Computing Joint Undertaking (EuroHPC JU)~\footnote{\href{https://eurohpc-ju.europa.eu/}{eurohpc-ju.europa.eu}} for funding this research by providing computing resources of the HPC system Vega at the Institute of Information sciences~\footnote{\href{https://www.izum.si/en/home/}{www.izum.si}}.

%\end{document}

%%%%%%%%%%%%%%%%%%%%%%%%%%%%%%%%%%%%%%%%%%%%%%%%%%%%%%%%%%%%%
%%%%%%%%%%%%%%%%%%%%%%%%%%%%%%%%%%%%%%%%%%%%%%%%%%%%%%%%%%%%%
%%%%%%%%%%%%%%%%%%%%%%%%%%%%%%%%%%%%%%%%%%%%%%%%%%%%%%%%%%%%%

\appendix

\begin{figure}[!b]
\centering
\includegraphics[width=0.97\columnwidth]{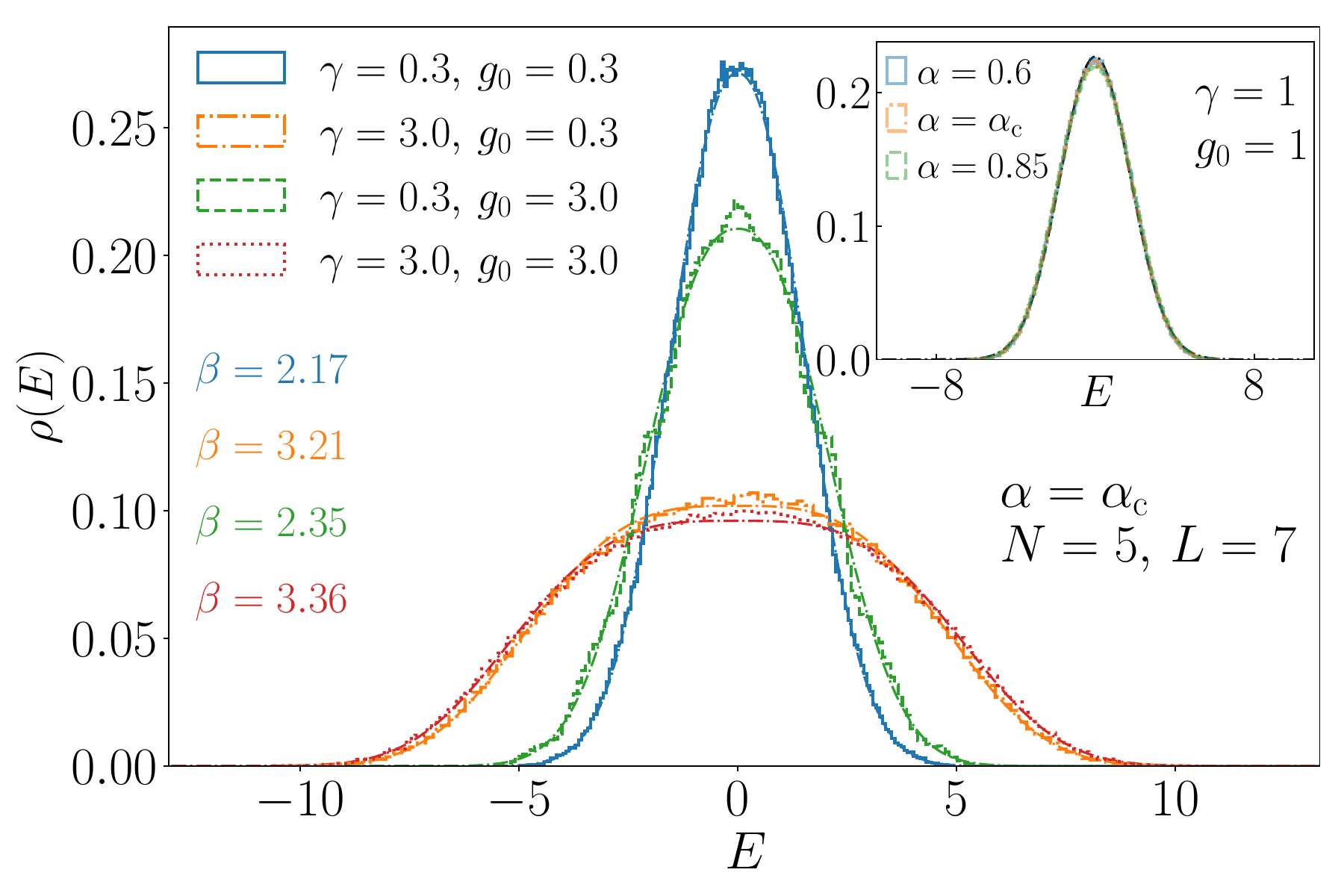}
\caption{
Density of states $\rho(E)$ in the QSM. 
The inset shows results at $g_0=\gamma=1$ and different values of $\alpha$.
Results are nearly indistinguishable from a normal distribution.
The main panel shows the results at $\alpha=\alpha_{\rm c}=1/\sqrt{2}$, when $g_0$ and $\gamma$ are considerably away from 1.
The dashed-dotted lines of the matching color show fits to the results using a generalized normal distribution $\rho_{\rm gen}(E)$ from Eq.~\eqref{eq:gen_normalized}. The values of the $\beta$ parameter of $\rho_{\rm gen}(E)$ are given in the legend, where values close to $\beta=2$ indicate proximity to the normal distribution. The results shown above were averaged over $N_\mathrm{samples}=40$ disorder realizations.
}
\label{fig_dos_qsun}
\end{figure}

\begin{figure*}[!]
\centering
\includegraphics[width=1\textwidth]{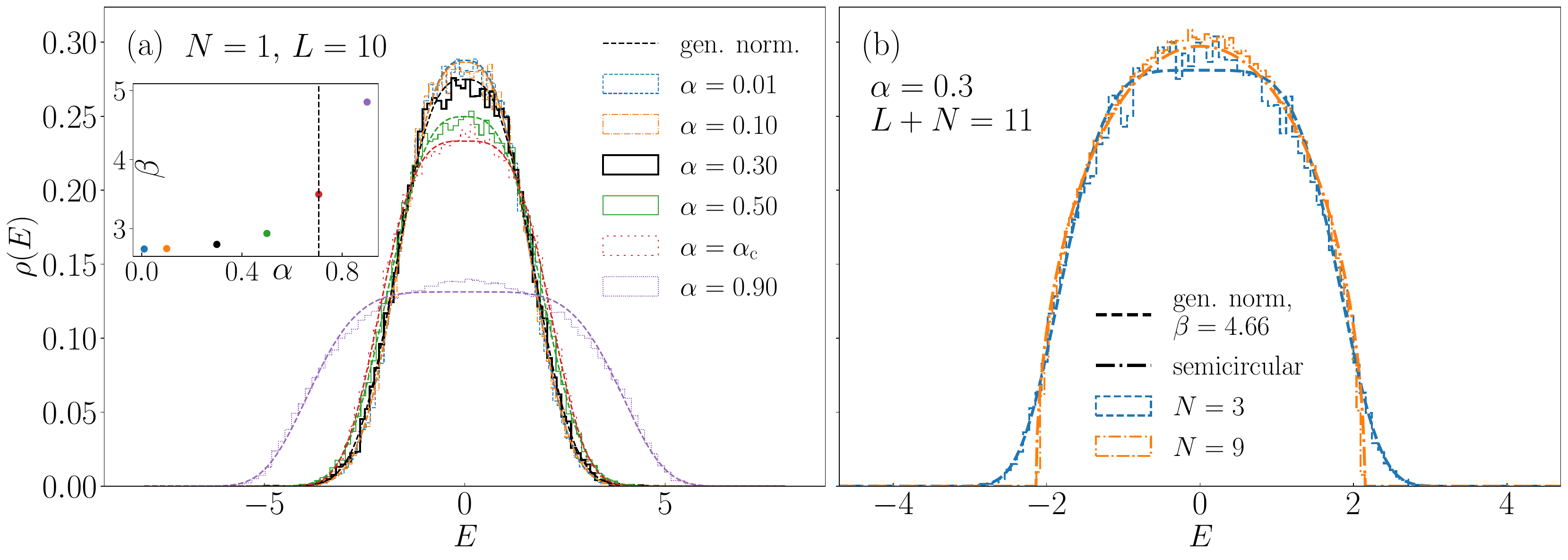}
\caption{
Density of states $\rho(E)$ in the UM at $J=1$ and $N+L=11$.
(a) $N=1$ at different values of $\alpha.$
Histograms show numerically obtained $\rho(E)$ while dashed lines of the matching color are fits according to the generalized normal distribution given by Eq.~\eqref{eq:gen_normalized}.
The inset shows the dependence of the $\beta$ parameter in Eq.~\eqref{eq:gen_normalized} on the value of the coupling parameter $\alpha$, with the vertical dotted line located at $\alpha=\alpha_{\rm c} = 1/\sqrt{2}$.
Note that even for small $\alpha,$ the distributions are not perfectly Gaussian, since $\beta \approx 2.77.$
(b) $\alpha=0.3$ at $N=3$ and $N=9$. 
At $N=3$, the numerical results are rather well described by a generalized normal distribution with $\beta\approx 4.66$, see the dashed line, while for $N=9,$ the numerics match with a Wigner semicircle distribution given by Eq.~\eqref{eq:wigner_semi}, see the dashed-dotted line. In both panels, the results shown were obtained by averaging over $N_{\mathrm{samples}}=20$ disorder realizations.
}
\label{fig_dos}
\end{figure*}

\section{Density of states} \label{sec:appendix_dos}

In Sec.~\ref{sec:models} we introduced both models under investigation, the QSM and the UM.
Here we provide additional information about the models by studying their coarse-grained density of states $\rho(E) = \delta N/\delta E$, which counts the number of Hamiltonian eigenstates $\delta N$ in a narrow energy window of width $\delta E$.

Figure~\ref{fig_dos_qsun} shows the density of states $\rho(E)$ in the QSM.
It is accurately described by the normal distribution when $g_0=\gamma=1$, see the inset of Fig.~\ref{fig_dos_qsun}, while deviations are observed if $\gamma$ and $g_0$ depart from 1, see the main panel of Fig.~\ref{fig_dos_qsun}.
In the latter case, we apply a phenomenological description of the density of states using the generalized normal distribution, 
\begin{equation}\label{eq:gen_normalized}
    \rho_{\rm gen}(E) = \frac{\beta}{2\sigma\Gamma(1/\beta)}\exp\left(-\left(|E-\mu|/\sigma\right)^\beta\right),
\end{equation}
where $E$ is the energy and $\Gamma$ is the Gamma function, while $\mu$ and $\sigma$ are the mean and the standard deviation of the energy distribution, respectively. We determine the parameter $\beta$ numerically and it controls the peakedness of the distribution with respect to its tails. For $\beta=2,$ one obtains the standard normal distribution, while the limiting case for $\beta\to\infty$ is the uniform distribution.
The variance $\sigma^2$ of the energy distribution can be estimated analytically as
\begin{equation} \label{eq:norm}
\begin{split}
 \Gamma^2_0 & = \langle \hat H^2 \rangle - \langle \hat H \rangle^2 = \\
 & = \gamma^2 + \frac{g_0^2}{16}\frac{1 - \alpha^{2L}}{1 - \alpha^2} + \frac{L}{4}\left(W^2 + \frac{\delta_W^2}{3}\right)\;,
\end{split}
\end{equation}
suggesting that at $\alpha < 1$, it is extensive ($\Gamma^2\propto L$) by the virtue of the last term in Eq.~(\ref{eq:def_model}).

The density of states $\rho(E)$ in the UM is for different parameter regimes shown in Fig.~\ref{fig_dos}.
Results at $N=1$ in Fig.~\ref{fig_dos}(a) show that $\rho(E)$ is well described by the generalized normal distribution $\rho_{\rm gen}(E)$ from Eq.~(\ref{eq:gen_normalized}) with $\beta \gtrsim 3$, i.e., it is not accurately described by the Gaussian distribution, which would correspond to $\beta=2$.
The reason for the deviation from Gaussianity is the RMT nature of the model, which includes processes beyond two-body terms.
The distribution is closest to the Gaussian at small $\alpha$, for which the suppression of the multi-body terms is the most efficient.
When $N$ is increased, $\rho(E)$ approaches the Wigner semicircle distribution~\cite{mehta_91}, given by
\begin{equation}\label{eq:wigner_semi}
    \rho_{\rm semi}(E) = \frac{2}{\pi R^2}\sqrt{R^2 - E^2}\;,
\end{equation}
with $R \approx 2$.
This is demonstrated for the case $N=9$ in Fig.~\ref{fig_dos}(b).

\begin{figure*}[!t]
\centering
\includegraphics[width=0.95\textwidth]{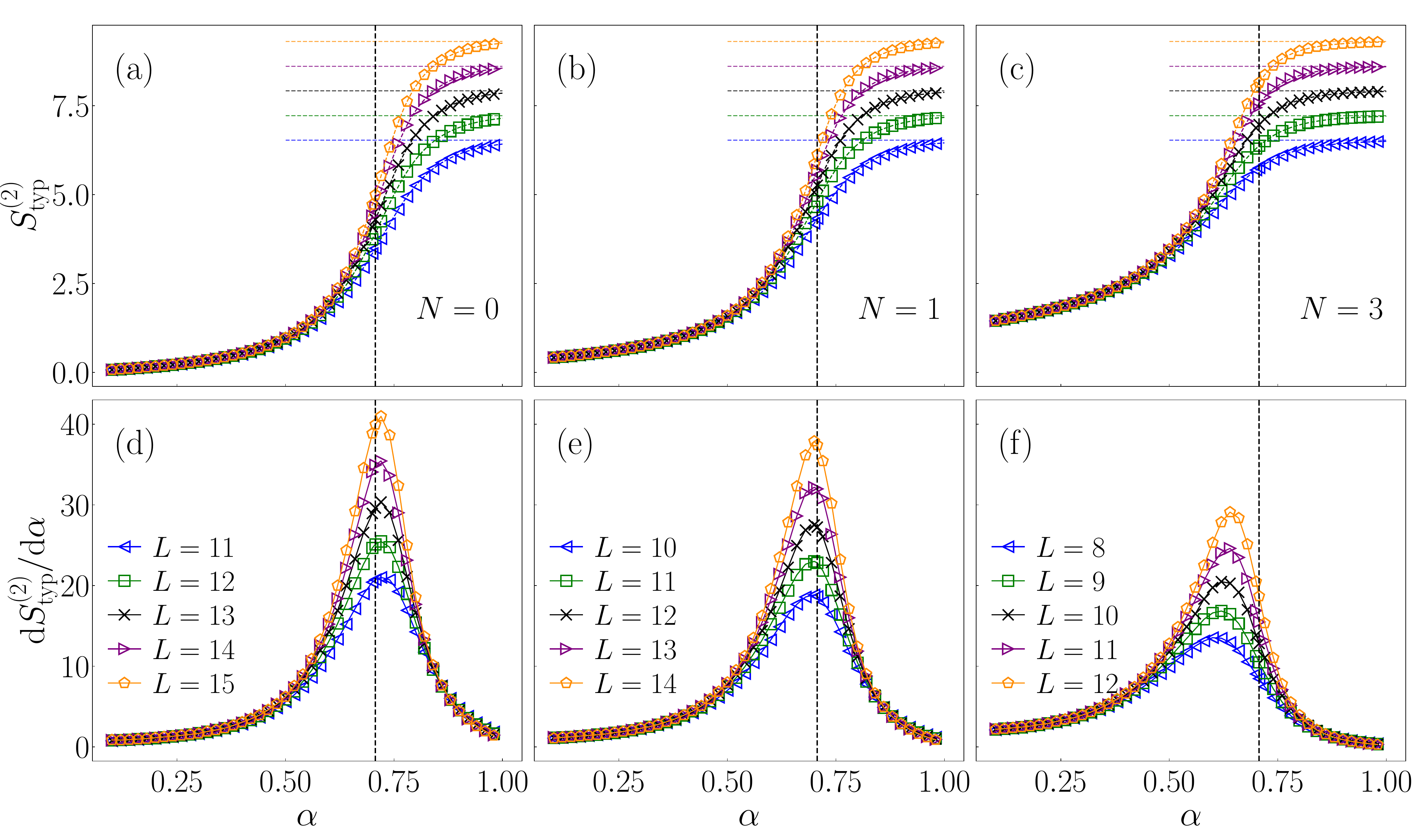}
\caption{
Participation entropies $S_{\rm typ}^{(2)}$ [panels (a-c)] and their derivatives $dS_{\rm typ}^{(2)}/d\alpha$ [panels (d-f)] in the UM at $J=1$.
(a) and (d) $N=0$,
(b) and (e) $N=1$ [the same results as in the main panels of Figs.~\ref{fig_sp2_toy}(c) and~\ref{fig_sp2_toy}(f), respectively],
(c) and (f) $N=3$.
Vertical dashed lines denote $\alpha=\alpha_c$ from Eq.~(\ref{eq:alpha_c}).
Results for $S_{\rm typ}^{(2)}$ are averaged over $N_\mathrm{samples}=1000,\, 400, \,300$ Hamiltonian realizations for $L + N\leq 13\,, L + N=14, \, L + N=15, $ respectively.
Averaging over the eigenstates, $\langle \cdots \rangle_n,$ was performed over $N_{\mathrm{eig}} = 1000$ eigenstates near the center of the spectrum for each data point.
}
\label{fig_sp1_toy}
\end{figure*}

\section{Further results for the participation entropy} \label{sec:appendix_Sp}

In Sec.~\ref{sec:ipr} we studied the participation entropies $S_{\rm typ}^{(q)}$ and their derivatives in both models.
In particular, in the UM we focused on systems with $N=1$, see Fig.~\ref{fig_sp2_toy}.
We complement these results in Fig.~\ref{fig_sp1_toy}, in which we also consider $S_{\rm typ}^{(q)}$ and their derivatives at $q=2$ in the systems with $N=0$, see Figs.~\ref{fig_sp1_toy}(a) and~\ref{fig_sp1_toy}(d), and $N=3$, see Figs.~\ref{fig_sp1_toy}(c) and~\ref{fig_sp1_toy}(f).
We focus on $q=2$ since the results in Fig.~\ref{fig_sp2_toy}(f) suggest that the derivatives of the participation entropies at large $q \gtrsim 2$ provide a good estimate of the critical point.

A general observation from Fig.~\ref{fig_sp1_toy} is that in the UM, the numerical results for all values of $N$ in the interval $0 \leq N \leq 3$ provide convincing evidence for the emergence of a critical point at $\alpha=\alpha_c$ from Eq.~(\ref{eq:alpha_c}), see the vertical dashed lines in Fig.~\ref{fig_sp1_toy}.
Nevertheless, one of the goals of this work is to focus on the parameter regimes in which the location of the critical point agrees with prediction from Eq.~(\ref{eq:alpha_c}) as accurately as possible already for rather small system sizes of the order of 10 spin-1/2 particles.
Then, Fig.~\ref{fig_sp1_toy}(e) suggests that this goal is most feasible at $N=1$, for which the peak of ${\rm d}S_{\rm typ}^{(q)}/{\rm d} \alpha$ almost exactly emerges at $\alpha=\alpha_c$ already in the system that consists of $N+L=11$ particles.
For this reason, the majority of the numerical studies of the UM was carried out for $N=1$.

\section{Further results for the entanglement entropy} \label{sec:appendix_Sr}

In Sec.~\ref{sec:Renyi} we studied the Rényi eigenstate entanglement entropies ${\cal S}^{(q)}$ and their derivatives in both models.
We focused on the case $p=1$, i.e., on entanglement entropies of subsystems that consist of a single spin-1/2 particle, which is most distant from (has the weakest coupling to) the ergodic quantum dot.

At $p=1$, the reduced density matrix describes a two-level system of a single spin-1/2 particle and hence it only contains two eigenvalues $\lambda_1$ and $\lambda_2$.
These two eigenvalues are correlated by the norm ${\rm Tr}\{\hat \rho_B\} = 1$, implying $\lambda_2 = 1 - \lambda_1$, and the Rényi entanglement entropy is then
\begin{equation} \label{eq:S_p1}
    {\cal S}^{(q)} = \frac{1}{\ln 2} \frac{1}{1-q} \Big\langle \Big\langle \ln(\lambda_1^q + (1-\lambda_1)^q) \Big\rangle_n \Big\rangle_H \;.
\end{equation}
This expression contains two limits.
In the limit $q\to 0$, one obtains [omitting the indices $n$ and $H$ in $\langle\cdots\rangle$]
\begin{equation} \label{eq:Sq_small}
    {\cal S}^{(q)} = 1 + \left[ 1 + \frac{\langle\langle\log_2(\lambda_1) + \log_2(1-\lambda_1)\rangle\rangle}{2} \right] q + O(q^2) \;,
\end{equation}
which suggest that ${\cal S}^{(q\to0)} \to 1$, as long as $\lambda_1$ is sufficiently away from 1.
This is expected to be the case in the ergodic phase and in the vicinity of the critical point.
In the limit $q \to \infty$, one can simplify $\langle\langle \ln(\lambda_1^q + (1-\lambda_1)^q) \rangle\rangle$ in Eq.~(\ref{eq:S_p1}) by replacing it with $q \langle\langle \ln \lambda_1 \rangle\rangle$, which is a reasonable approximation if $\lambda_1$ is sufficiently larger than $1/2$.
This is a natural assumption for the nonergodic phase and, as shown in Sec.~\ref{sec:Schmidt}, also in the vicinity of the critical point.
It then follows that the leading term at $q \to\infty$ is
\begin{equation} \label{eq:Sq_large}
    {\cal S}^{(q\to\infty)} \approx -\log_2 \lambda_1 \;.
\end{equation}
This result suggests that, as expected, the two-level system is maximally entangled at the critical point if $\lambda_1 \approx 1/2$, while the entanglement is vanishingly small if $\lambda_1 \to 1$.
In Sec.~\ref{sec:Schmidt} we showed that, at least for the model parameters under investigation, $\lambda_1$ at the critical point is neither close to $1/2$ nor to $1$, which gives rise to substantial, but not maximal entanglement of ${\cal S}^{(q\to\infty)}$.
Results for the entanglement entropies at $p=1$, shown in Figs.~\ref{fig_sr1_toy} and~\ref{fig_sr1_sun}, are consistent with these considerations.

\begin{figure*}[!t]
\centering
\includegraphics[width=0.77\textwidth]{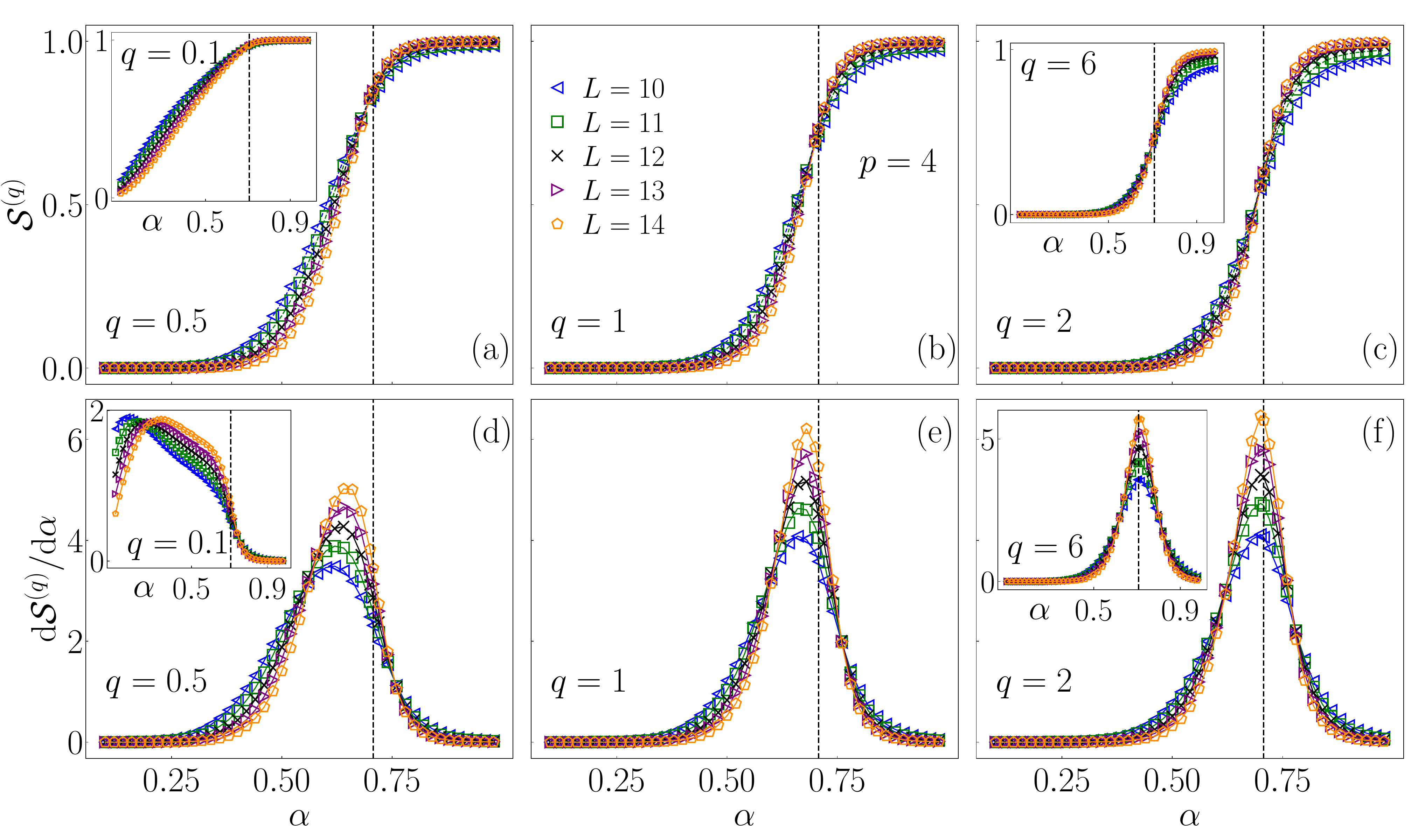}
\caption{
The Rényi entanglement entropies ${\cal S}^{(q)}$ [panels (a-c)] and their derivatives ${\rm d}{\cal S}^{(q)}/{\rm d}\alpha$ [panels (d-f)] in the UM at $J=1$, $N=1$ and $p=4$.
(a) and (d) $q=0.5$ in the main panel and $q=0.1$ in the insets.
(b) and (e) $q=1$.
(c) and (f) $q=2$ in the main panel and $q=6$ in the inset.
Vertical dashed lines denote $\alpha=\alpha_{\rm c}$ from Eq.~(\ref{eq:alpha_c}).
Results for ${\cal S}^{(q)}$ are averaged over $N_\mathrm{samples}=1000,\, 400, \,300$ Hamiltonian realizations for $L\leq 12\,, L=13, \, L=14, $ respectively. 
Averaging over the eigenstates, $\langle \cdots \rangle_n,$ was performed over $N_{\mathrm{eig}} = 1000$ eigenstates near the center of the spectrum for each data point.
}
\label{fig_sr2_toy}
\end{figure*}

\begin{figure*}[!t]
\centering
\includegraphics[width=0.77\textwidth]{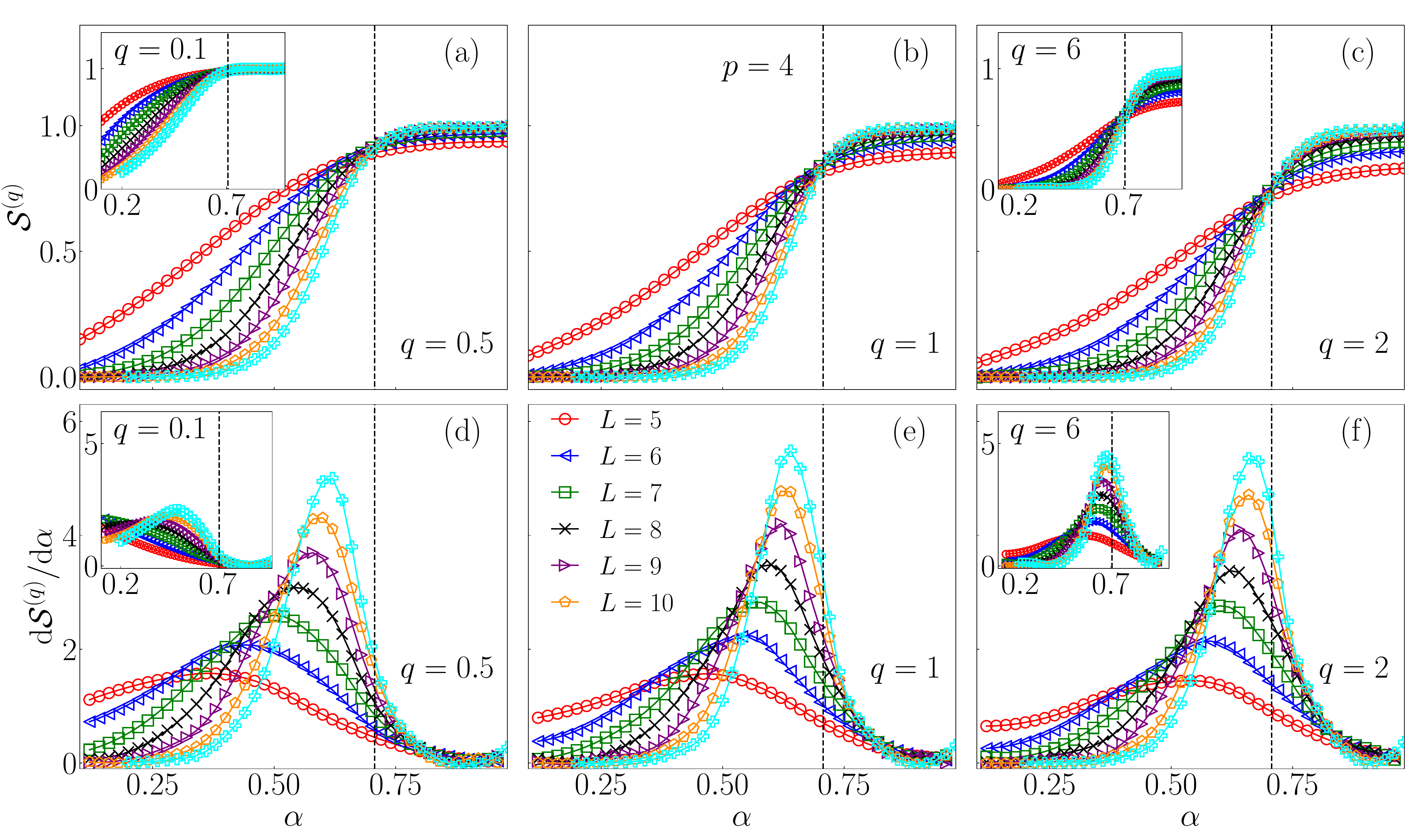}
\caption{
The Rényi entanglement entropies ${\cal S}^{(q)}$ [panels (a-c)] and their derivatives ${\rm d}{\cal S}^{(q)}/{\rm d}\alpha$ [panels (d-f)] in the QSM at $g_0=\gamma=1$, $N=5$ and $p=4$.
(a) and (d) $q=0.5$ in the main panel and $q=0.1$ in the insets.
(b) and (e) $q=1$.
(c) and (f) $q=2$ in the main panel and $q=6$ in the inset.
Vertical dashed lines denote $\alpha=\alpha_{\rm c}$ from Eq.~(\ref{eq:alpha_c}).
Results for ${\cal S}^{(q)}$ are averaged over $N_\mathrm{samples}=500$ Hamiltonian realizations.
Averaging over the eigenstates, $\langle \cdots \rangle_n,$ was performed over $N_{\mathrm{eig}} = 500$ eigenstates near the center of the spectrum for each data point.
}
\label{fig_sr2_sun}
\end{figure*}

The above analysis is in Figs.~\ref{fig_sr2_toy} and~\ref{fig_sr2_sun} extended to the case $p=4$, i.e., to subsystems that consist of four most distant spin-1/2 particles.
For both models, the results in Figs.~\ref{fig_sr2_toy} and~\ref{fig_sr2_sun} suggest that the critical point can be rather accurately determined from the scale invariant point of ${\cal S}^{(q)}$ emerging at $\alpha=\alpha_c$ as well as from the peak in its derivative ${\rm d}{\cal S}^{(q)}/{\rm d}\alpha$.

Being more quantitative, we note that in the case of the UM, we actually observe no major differences between the $p=1$ and $p=4$ cases, since the results in Figs.~\ref{fig_sr1_toy} and~\ref{fig_sr2_toy} are virtually almost indistinguishable.
In the QSM, however, some differences can be observed between the $p=1$ case in Fig.~\ref{fig_sr1_sun} and the $p=4$ case in Fig.~\ref{fig_sr2_sun}.
In particular, at $p=4$ the crossing point of ${\cal S}^{(q)}$ versus $\alpha$ at small system sizes $N+L \approx 10$ emerges slightly away from $\alpha=\alpha_c$, and it drifts towards $\alpha_c$ upon increasing $L$.
This effect is especially apparent at small values of $q$ shown in Fig.~\ref{fig_sr2_sun}(a).
These results establish the view that the sharpest signatures of the critical point are in the QSM contained in properties of the particles that are most distant from the ergodic quantum dot.

\section{Further results for the Schmidt gap} \label{sec:appendix_Sgap}

In the insets of Figs.~\ref{fig_Sgap_toy} and~\ref{fig_Sgap_sun} in Sec.~\ref{sec:Schmidt} we plotted the second Rényi entanglement entropies $\mathcal{S}^{(2)}$ as functions of the Schmidt gap $\Delta$ and observed a nearly perfect collapse of the results for all studied system sizes and for both models.
The quantities $\mathcal{S}^{(2)}$ and $\Delta$ were defined in Eqs.~(\ref{eq:def_renyi}) and~(\ref{eq:def_Sgap}), respectively, as averages over Hamiltonian eigenstates and different Hamiltonian realizations.

\begin{figure}[!t]
\centering
\includegraphics[width=1\columnwidth]{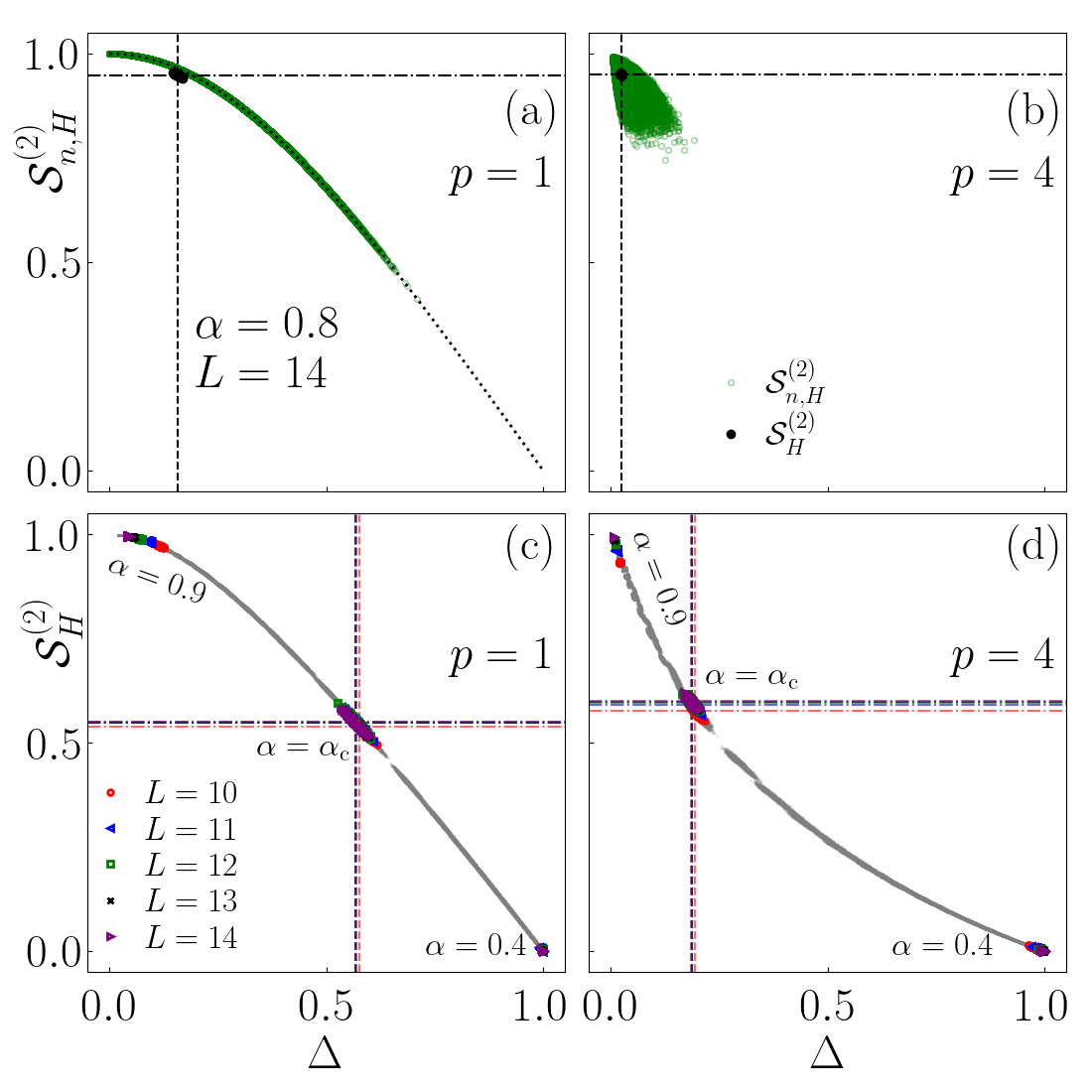}
\caption{
The role of averaging on the dependence of the Rényi entanglement entropies $\mathcal{S}^{(2)}$ on the Schmidt gap $\Delta$ in the UM at $J=1$ and $N=1$.
(a) $p=1$ and (b) $p=4$, both at $\alpha=0.8$ and $L=14$.
The small green symbols denote ${\mathcal S}^{(2)}_{n,H}$ versus $\Delta_{n,H}$, see Eq.~(\ref{eq:def_renyi_n_H}), while the larger black symbols denote ${\mathcal S}^{(2)}_{H}$ versus $\Delta_{H}$, see Eq.~(\ref{eq:def_renyi_H}).
The dotted line in (a) is the result from Eq.~(\ref{eq:Sq_vs_Delta}),
while the horizontal and vertical lines in (a) and (b) denote the averages $\mathcal{S}^{(2)}$ and $\Delta$ from Eqs.~(\ref{eq:def_renyi}) and~(\ref{eq:def_Sgap}), respectively, at $\alpha=0.8$.
In panels (c) and (d) we then extend the analysis of ${\mathcal S}^{(2)}_{H}$ versus $\Delta_{H}$ to $\alpha$ in the interval $\alpha\in[0.1, 0.98]$ and to different values of $L$, as indicated in the legend in (c).
Results for most values of $\alpha$ are shown as gray symbols.
The colored symbols specifically show the results for different system sizes at three different values of $\alpha=0.9$, $\alpha=\alpha_c$ and $\alpha=0.4$.
The horizontal and vertical lines denote the averages $\mathcal{S}^{(2)}$ and $\Delta$ at a given $L$ at $\alpha=\alpha_c$.
The results in panels (a) and (b) were obtained using averaging over $N_{\mathrm{samples}}=300$ Hamiltonian realizations.
Results in panels (c) and (d) were obtained by averaging over $N_\mathrm{samples}=1000, \, 400, \, 300$ for $L\leq 12, \, L=13, \, L=14,$ respectively.}
\label{fig_sGap_toy_2}
\end{figure}

To better understand the origin of the emergence of $\mathcal{S}^{(2)}$ being a well-defined function of $\Delta$, we here study both quantities before the averages are carried out.
Specifically, we define the entanglement entropy ${\mathcal S}^{(q)}_{n,H}$ and the Schmidt gap $\Delta_{n,H}$ of a single eigenstate $|n\rangle$ of a single Hamiltonian realization as
\begin{equation}\label{eq:def_renyi_n_H}
    {\mathcal S}^{(q)}_{n,H} = \frac{1}{\ln{\mathcal{D}_B}} \frac{1}{1 - q} 
    \ln \sum\limits_{i=1}^{\mathcal{D}_B} \lambda_i^q \;, \;\;\;
    \Delta_{n,H} = \lambda_1 - \lambda_2 \;.
\end{equation}
Analogously, we define the entanglement entropy ${\mathcal S}^{(q)}_{H}$ and the Schmidt gap $\Delta_{H}$ of an average over Hamiltonian eigenstates of a single Hamiltonian realization as
\begin{equation}\label{eq:def_renyi_H}
    {\mathcal S}^{(q)}_H = \frac{1}{\ln{\mathcal{D}_B}} \frac{1}{1 - q} \Big\langle \ln \sum\limits_{i=1}^{\mathcal{D}_B} \lambda_i^q \Big\rangle_n\;, \;\;\;
    \Delta_{H} = \Big\langle \lambda_1 - \lambda_2 \Big\rangle_n \;.
\end{equation}
In Fig.~\ref{fig_sGap_toy_2} we study the properties of the quantities defined in Eqs.~(\ref{eq:def_renyi_n_H}) and~(\ref{eq:def_renyi_H}) in the UM at $J=1$ and $N=1$.

Figure~\ref{fig_sGap_toy_2}(a) shows both ${\mathcal S}^{(2)}_{n,H}$ versus $\Delta_{n,H}$ and ${\mathcal S}^{(2)}_{H}$ versus $\Delta_{H}$ at $p=1$ and $\alpha=0.8$.
As expected from Eq.~(\ref{eq:Sq_vs_Delta}), the entanglement entropy at $p=1$ is a unique function of Schmidt gap already on a level of a single eigenstate.
On the other hand, this is not the case at $p=4$, see Fig.~\ref{fig_sGap_toy_2}(b), in which the results for ${\mathcal S}^{(2)}_{n,H}$ versus $\Delta_{n,H}$ give rise to a wide cloud of points without any well defined functional dependence.

The next question that we then ask is at which level of averaging the second Rényi entanglement entropy becomes a well-defined function of the Schmidt gap even at $p=4$, as suggested by the insets of Figs.~\ref{fig_Sgap_toy}(b) and~\ref{fig_Sgap_sun}(b).
In Figs.~\ref{fig_sGap_toy_2}(c) and~\ref{fig_sGap_toy_2}(d) we plot ${\mathcal S}^{(2)}_{H}$ versus $\Delta_{H}$ at $p=1$ and $p=4$, respectively.
The results suggest that at $p=4$, the averaging over Hamiltonian eigenstates within a single Hamiltonian realization represents the key contribution to establishing a well-defined functional dependence of ${\mathcal S}^{(2)}$ versus $\Delta$.
Still, a careful inspection of Fig.~\ref{fig_sGap_toy_2}(d) reveals that the collapse of the data to a single function is not perfect, and hence the functional dependence of ${\mathcal S}^{(2)}$ versus $\Delta$ at $p>1$ should not be considered as an exact property.

\section{Further results for the fractal dimension} \label{sec:appendix_Fractal}

In Fig.~\ref{fig_tauq1_sun} of the main text, we showed the results for the participation entropy $S_{\rm typ}^{(q)}$ versus the logarithm of the Fock space dimension $\ln{\cal D}$ for the QSM.
They were accurately described by the functional ansatz from Eq.~(\ref{eq:def_Sq_dq}) at the critical point as well as deep in the ergodic and nonergodic phases.
Using this ansatz, we obtained the fractal dimension $d_{\rm typ}^{(q)}$ studied in Sec.~\ref{sec:multifractality}.

\begin{figure}[!t]
\centering
\includegraphics[width=1\columnwidth]{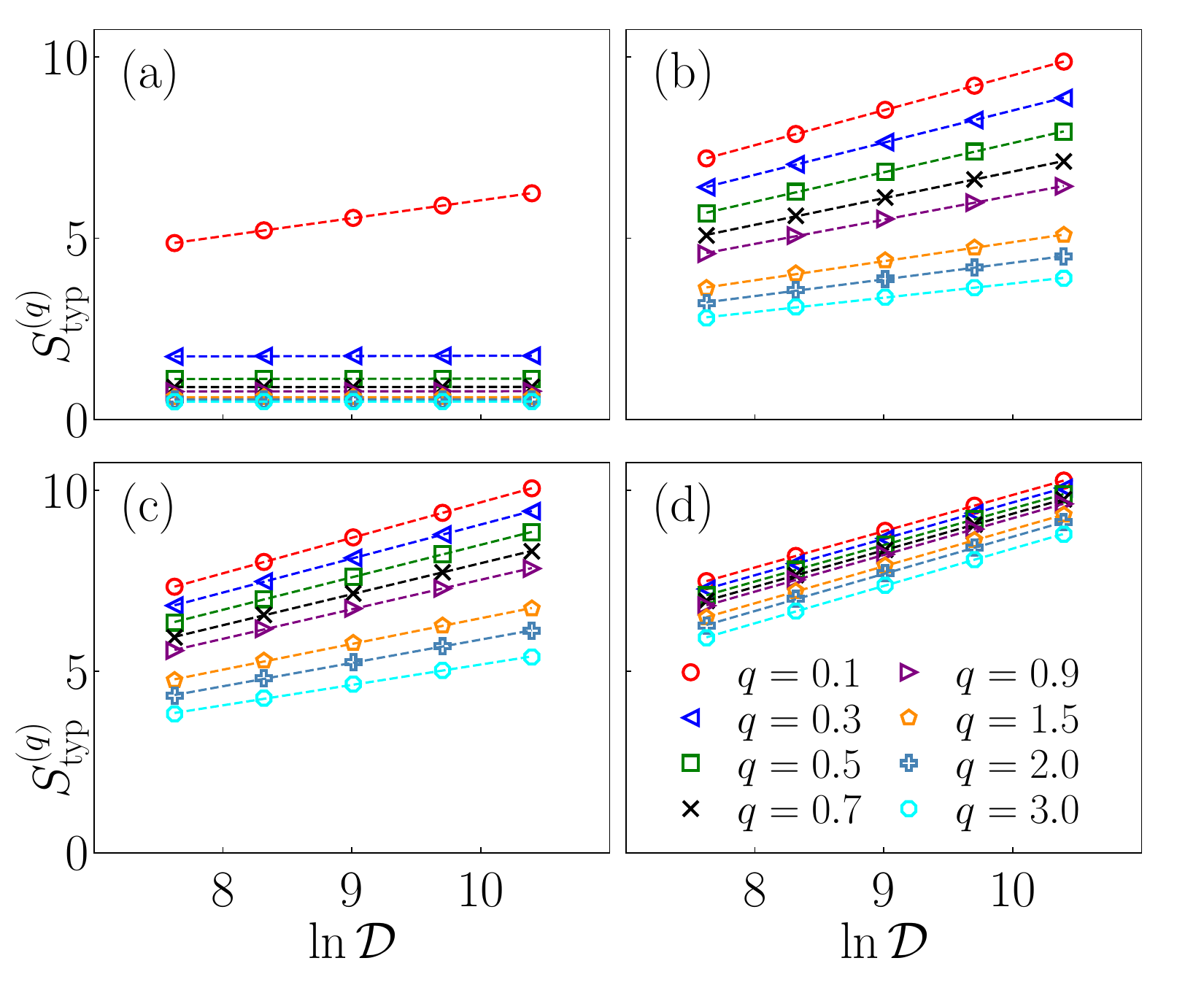}
\caption{
Participation entropy $S_{\rm typ}^{(q)}$ versus the logarithm of the Fock space dimension $\ln {\cal D}$ in the UM at $N=1$ and: (a) $J=1,  \, \alpha=0.2,$ (b) $J=0.5, \, \alpha=\alpha_c,$  (c) $J=1, \, \alpha=\alpha_c,$ (d) $J=1, \, \alpha=0.9$.
Lines are fits according to the ansatz in Eq.~(\ref{eq:def_Sq_dq}) for Fock space dimensions of the total system in the range from ${\cal D}=2^{11}$ to ${\cal D}=2^{15}$.
They allow for the extraction of the fractal dimension $d^{(q)}_\mathrm{typ}$ and the coefficient $\tau^{(q)}_\mathrm{typ}$, plotted in Fig.~\ref{fig_tauq1_toy_short}.
}
\label{fig_tauq1_toy}
\end{figure}

Analogous results for the UM are shown in Fig.~\ref{fig_tauq1_toy}.
They are also accurately described by the functional ansatz from Eq.~(\ref{eq:def_Sq_dq}), i.e., the system is said to be very close to the asymptotic regime in which the $L$ dependence of the fractal dimension $d_{\rm typ}^{(q)}$ is likely very small.
The only exception may be the case for $q=0.1$, see Figs.~\ref{fig_tauq1_toy}(a), in which $S_{\rm typ}^{(q)}$ still increases with $\ln{\cal D}$ even though at larger values of $q$ we observe no increase.

%\clearpage
\bibliographystyle{biblev1}
\bibliography{references1,references2,refsWojciech}

\end{document}